\pdfoutput=1
\documentclass[12pt,preprint]{aastex}
\shorttitle{Reddening toward the Galactic bulge}
\shortauthors{Nataf et al.}
\usepackage{subfigure}
\usepackage{array,longtable}
\usepackage{natbib}
\usepackage{float}
\usepackage{sidecap}
\usepackage{amsmath}

\begin{document}
\title{Reddening and Extinction Toward the Galactic Bulge from OGLE-III: The Inner Milky Way's R$_{V}\sim2.5$ Extinction Curve\footnote{Based on observations obtained with the 1.3 m Warsaw telescope at the Las Campanas Observatory of the Carnegie Institution for Science.}}
\author{David M. Nataf\altaffilmark{1}, Andrew Gould\altaffilmark{1}, Pascal Fouqu\'e\altaffilmark{2,3},  Oscar A. Gonzalez\altaffilmark{4}, Jennifer A. Johnson\altaffilmark{1},  Jan Skowron\altaffilmark{1}, Andrzej Udalski\altaffilmark{5}, Micha{\l} K. Szyma{\'n}ski\altaffilmark{5}, Marcin Kubiak\altaffilmark{5}, Grzegorz Pietrzy{\'n}ski\altaffilmark{5,6}, Igor Soszy{\'n}ski\altaffilmark{5}, Krzysztof Ulaczyk\altaffilmark{5}, {\L}ukasz Wyrzykowski\altaffilmark{5,7}, Rados{\l}aw Poleski\altaffilmark{5} }
\altaffiltext{1}{Department of Astronomy, Ohio State University, 140 W. 18th Ave., Columbus, OH 43210}
\altaffiltext{2}{Universit\'e de Toulouse; UPS-OMP; IRAP; Toulouse, France}
\altaffiltext{3}{CNRS; IRAP; 14, avenue Edouard Belin, F-31400 Toulouse, France}
\altaffiltext{4}{European Southern Observatory, Karl-Schwarzschild-Strasse 2, D-85748 Garching, Germany}
\altaffiltext{5}{Warsaw University Observatory, Al. Ujazdowskie 4, 00-478 Warszawa,Poland}
\altaffiltext{6}{Universidad de Concepci{\'o}n, Departamento de Astronomia,
Casilla 160--C, Concepci{\'o}n, Chile}
\altaffiltext{7}{Institute of Astronomy, University of Cambridge, Madingley Road, Cambridge CB3 0HA, UK}
\email{nataf@astronomy.ohio-state.edu}

\begin{abstract}
We combine $VI$ photometry from OGLE-III with $VVV$ and 2MASS measurements of $E(J-K_{s})$ to resolve the longstanding problem of the non-standard optical extinction toward the Galactic bulge. We show that the extinction is well-fit by the relation $A_{I} = 0.7465{\times}E(V-I) + 1.3700{\times}E(J-K_{s})$, or, equivalently, $A_{I} = 1.217{\times}E(V-I)(1+1.126{\times}(E(J-K_{s})/E(V-I)-0.3433))$.  The optical and near-IR reddening law toward the inner Galaxy approximately follows an $R_{V} \approx 2.5$ extinction curve with a dispersion ${\sigma}_{R_{V}} \approx 0.2$, consistent with extragalactic investigations of the hosts of type Ia SNe. Differential reddening is shown to be significant on scales as small as as our mean field size of 6$\arcmin$. The intrinsic luminosity parameters of the Galactic bulge red clump (RC) are derived to be $(M_{I,RC}, \sigma_{I,RC,0}, (V-I)_{RC,0}, \sigma_{(V-I)_{RC}}, (J-K_{s})_{RC,0}) = (-0.12, 0.09, 1.06, 0.121, \newline 0.66)$. Our measurements of the RC brightness, brightness dispersion and number counts allow us to estimate several Galactic bulge structural parameters. We estimate a distance to the Galactic center of 8.20 kpc. We measure an upper bound on the tilt $\alpha \approx 40^{\circ}$ between the bulge's major axis and the Sun-Galactic center line of sight, though our brightness peaks are consistent with predictions of an N-body model oriented at $\alpha \approx 25^{\circ}$. The number of RC stars suggests a total stellar mass for the Galactic bulge of $\sim2.3{\times}10^{10} M_{\odot}$ if one assumes a canonical Salpeter IMF, or  $\sim1.6{\times}10^{10} M_{\odot}$ if one assumes a bottom-light Zoccali IMF.
\end{abstract}

\section{Introduction}
\label{sec:Introduction}
The central bulge of the Milky Way Galaxy is the only stellar spheroid for which we can measure detailed abundances, ages and all six phase space dimensions for individual stars, as well as the luminosity function and spatial distribution for the population as a whole. Some $\sim$10\% of the Milky Way's stars are bulge stars, including a disproportionate number of the oldest and most metal-rich stars. It is therefore evident that any theory of Galaxy formation and evolution is required to reproduce the observed properties of the bulge, and conversely, that the properties of the bulge should be measured as precisely and accurately as possible to best discriminate between different Galaxy formation models. 

However, as scientifically desirable as this greater project may be, it is also difficult due to several challenges that prevent further, deeper understanding of the bulge\footnote{Henceforth, we almost exclusively refer to the bar/bulge of the Milky Way as the bulge for the sake of consistent representation. We do recognise that these two words have different meanings, but the kinematic decomposition of the Galaxy's central population remains a matter of active investigation and controversy at this time.}. There are significant correlations between chemistry and kinematics \citep{2010A&A...519A..77B,2012A&A...546A..57U,2012ApJ...756...22N}, distinct chemical subgroups \citep{2012arXiv1205.4715A}, gradients in metallicity \citep{2008A&A...486..177Z}, deviations from the classical picture of the triaxial ellipsoid at large separations from the minor axis \citep{2005ApJ...630L.149B,2008A&A...491..781C}, and both large  \citep{2010ApJ...721L..28N,2010ApJ...724.1491M} and small \citep{2001A&A...379L..44A,2005ApJ...621L.105N} separations from the plane. These issues necessitate larger data sets and better models. The viewing angle $\alpha$ between the bulge's major axis and the sun-Galactic center line of sight remains undetermined, with best-fit values ranging from from $\alpha=13^{\circ}$ \citep{2007A&A...465..825C}  to $\alpha=44^{\circ}$ \citep{2005ApJ...630L.149B}. This prevents further understanding of the inner Galaxy's gravitational potential, as uncertainties in the value of $\alpha$ are degenerate with those of the bulge's axis ratios \citep{1997ApJ...477..163S} and rotation speed \citep{2010ApJ...720L..72S}.

The most significant source of uncertainty, however, is the extinction. It averages $A_{K} \approx 3$ toward the  Galactic center \citep{2010A&A...511A..18S}, suggesting $A_{V} \approx 50$ \citep{2008ApJ...680.1174N}. For most of the bulge, values of $A_{V} = 2$ are typical \citep{2004MNRAS.349..193S}. The high values of reddening close to the plane render it quite difficult to obtain spectroscopic observations, proper motions, and stellar density maps. Further from the plane, these can be obtained, but not fully understood due to significant zero-point uncertainties in the extinction, and indirectly, the distance.

A further complication arises from the fact the extinction toward the inner Galaxy is not only large but also non-standard. This was first suggested by \citet{2000ApJ...528L...9P} as a solution to the anomalous colors of bulge RR Lyrae \citep{1999ApJ...521..206S} and red clump (RC) stars \citep{1999AcA....49..319P}.  \citet{2001ApJ...547..590G} and \citet{2003ApJ...590..284U} were the first to demonstrate that the reddening law toward the inner Galaxy is described by smaller total-to-selective ratios than the ``standard'' values measured for the local interstellar medium, implying a steeper extinction curve and thus a smaller characteristic size for dust grains \citep{2003ARA&A..41..241D}. \citet{2003ApJ...590..284U} measured $ dA_{I}/dE(V-I) \approx 1.1$ (denoted  ${\Delta}A_{I}/{\Delta}E(V-I)$ in that work) toward several bulge fields, much smaller than the value of $dA_{I}/dE(V-I) \approx 1.45$ suggested by the standard interstellar extinction curve of $R_{V}=3.1$ \citep{1989ApJ...345..245C,1994ApJ...422..158O}.  \citet{2003ApJ...590..284U} showed that applying the same methodology to observations of the Large Magellanic Cloud taken with the same instruments yielded $dA_{I}/dE(V-I) \approx 1.44$, the standard value, demonstrating the robustness of the result. Further, not only was the reddening law toward the bulge found to be non-standard, it was also found to be rapidly varying between sightlines, with values of $dA_{I}/dE(V-I)$ ranging from $0.94\pm0.01$ to $1.16\pm0.03$. The steeper extinction law toward the inner Galaxy has been subsequently confirmed with observations using \textit{Hubble Space Telescope (HST)} optical filters by \citet{2010A&A...515A..49R}, by analysis of RR Lyrae stars in OGLE-III \citep{2012ApJ...750..169P}, and also in the near-IR \citep{2008ApJ...680.1174N,2009MNRAS.394.2247G,2010A&A...511A..18S}. Meanwhile, \citet{2009ApJ...707..510Z} and \citet{2011ApJ...737...73F} both found that the extinction law toward the inner Galaxy was \textit{shallower} (greyer) in the mid-IR. 

The variations in both the extinction and the extinction law made it difficult to reliably trace the spatial structure of the bulge \citep{2010AcA....60...55M}. Applying the $VI$ extinction maps of \citet{2004MNRAS.349..193S} to the bulge color-magnitude diagram (CMD) implied a distance to the Galactic center of $\sim$9 kpc \citep{2007MNRAS.378.1064R,2009A&A...498...95V}, a large value relative to the geometrically determined distances to the Galactic center of $7.62 \pm 0.32$ kpc \citep{2005ApJ...628..246E}, $8.27 \pm 0.29$ kpc \citep{2012arXiv1207.3079S},  and $8.4 \pm 0.4$ kpc \citep{2008ApJ...689.1044G}. As the structure of the inner Galaxy is a very sensitive probe of the environmental conditions in which the Galaxy formed and evolved \citep{2005MNRAS.358.1477A,2012MNRAS.422.1902I}, an accurate spatial determination of the bulge's morphology would yield a powerful test of Galaxy formation models. Moreover, investigations of the metallicity distribution function of bulge giants have had to rely on imprecise and potentially inaccurate estimates of surface gravity and photometric temperature, further reducing our ability to probe the primordial conditions of the Galaxy.

We resolve these issues in this investigation by combining OGLE-III observations in $V$ and $I$ with $VVV$ and 2MASS measurements of $E(J-K_{s})$ \citep{2012A&A...543A..13G}. We confirm previous findings that the $VI$ extinction toward the inner Galaxy is steeper than standard, but also show that it is a little less steep than previously assumed. We show that this is likely due to an effect we label ``composite extinction bias'', which makes it unphysical to extrapolate a slope of $dA_{I}/dE(V-I)$ to $E(V-I)=0$. Our parameterization for the extinction,  $A_{I} = 0.7465{\times}E(V-I) + 1.3700{\times}E(J-K_{s})$, is less sensitive to the rapid variations in the extinction law than the computation of slopes $dA_{I}/dE(V-I)$. Whereas the latter must be computed from an ensemble of measurements spread across 30$\arcmin$ or more, the former can be directly measured for each $\sim 6\arcmin \times 6\arcmin$ sightline.

The structure of this paper is as follows. We summarize the data used in Section \ref{sec:Data}. Our methodology for measuring the parameters of the RC is described in Section \ref{sec:RCmeasurements}, and we derive  the intrinsic RC luminosity parameters in Section \ref{sec:Calibration}. We briefly state the properties of reddening that would be expected using a standard reddening curve in Section \ref{Sec:ReddeningTheory}. The reddening measurements are presented and discussed in Section \ref{sec:Reddening}, including comparisons to the reddening maps of Schlegel, Finkbeiner \& Davis (SFD, \citealt{1998ApJ...500..525S}) and the derivation of an empirical rule to estimate differential reddening. Methods to convert the reddening into an extinction using a single color are demonstrated to inevitably fail in Section \ref{sec:ReddeningLawEstimates}, and a more successful extinction law is derived in Section \ref{sec:RJKVI} by including information from both $E(V-I)$ and $E(J-K_{s})$. We demonstrate that reddening constraints from MACHO photometry may have been misinterpreted in Section \ref{sec:MACHO}. In section \ref{sec:GalacticStructure}, we show that our dereddened apparent magnitudes suggest a distance to the Galactic center $R_{0}=8.20$ kpc, and a tilt between the Galactic bulge's major axis and the sun-Galactic center line of sight no greater than $\alpha \approx 41^{\circ}$. We translate these measurements into constraints for microlensing events toward the bulge in Section \ref{sec:MicrolensingConstraint}. We analyze our number counts for the RC in Section \ref{sec:GalacticStructure2} and show that combining these with the assumptions of standard stellar evolution and a Salpeter IMF yields an estimated Galactic bulge stellar mass of $M \sim 2.3{\times}10^{10} M_{\odot}$. The thickness of the Galactic bulge is discussed in Section \ref{sec:GalacticStructure3}. The data structure is summarized in Section \ref{Sec:DataSummary}. Results are discussed in Section \ref{sec:Discussion}.

\section{Data}
\label{sec:Data}
\begin{figure}[H]
\begin{center}
\includegraphics[totalheight=0.4\textheight]{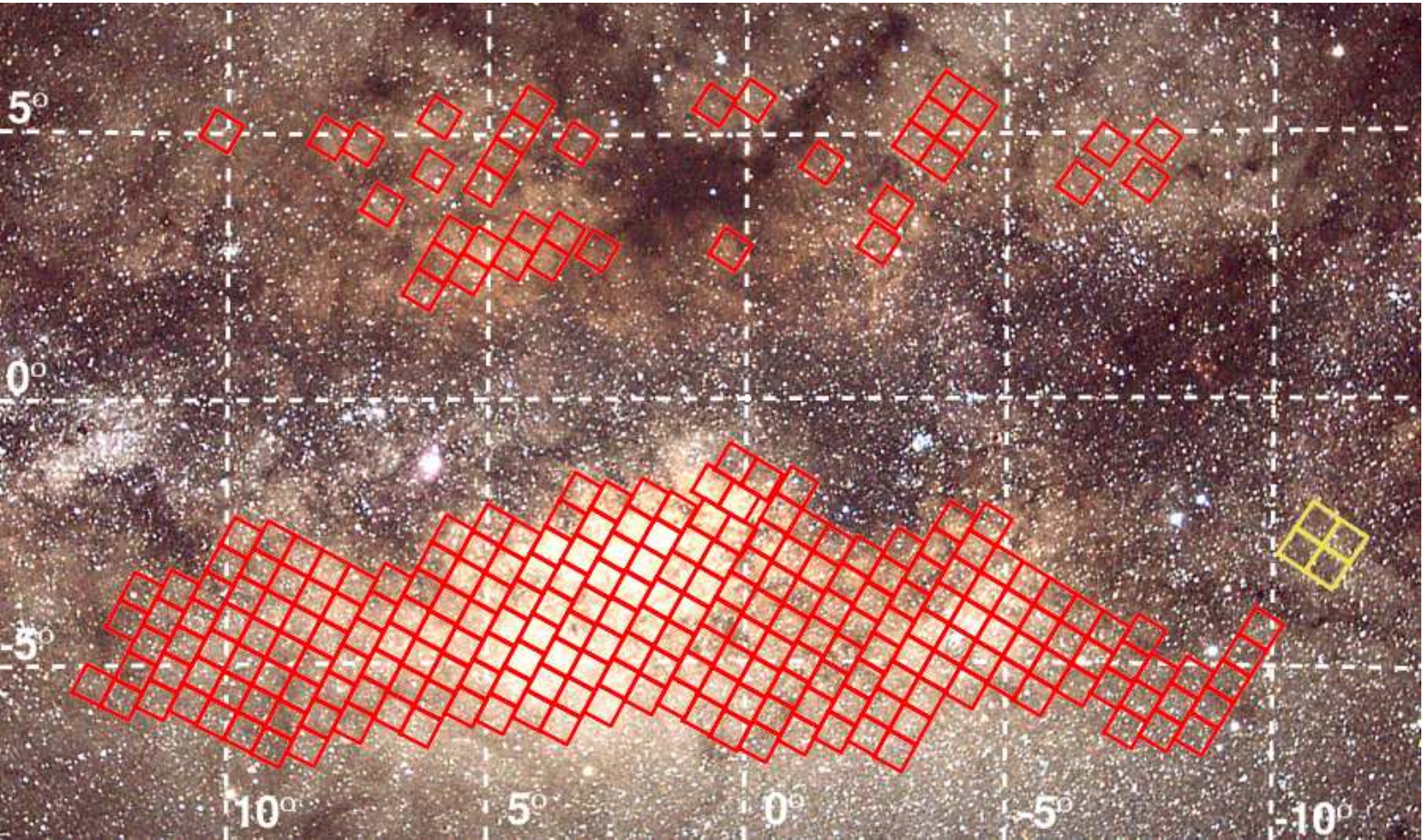}
\end{center}
\caption{Coverage of the OGLE-III Galactic bulge photometric survey used in this work, overplotted on an optical image of the same area. Galactic coordinate system shown. Red squares denote the OGLE-III fields used in this work, and yellow squares denote fields not used.  } 
\label{Fig:ogle3-blg-fields}
\end{figure}

OGLE-III observations were taken with the 1.3 meter Warsaw Telescope, located at the Las Campanas Observatory. The camera has eight 2048x4096 detectors, with a combined field of view of $0.6^{\circ}\times
0.6^{\circ}$ yielding a scale of approximately 0.26$\arcsec$/pixel. We use observations from 263 of the 267 OGLE-III fields directed toward the Galactic bulge, which are almost entirely within the range $-10^{\circ}<l<10^{\circ}$ and $2^{\circ}<|b|<7^{\circ}$. We do not use 4 of the fields, BLG200, 201, 202, and 203; located toward $(l,b) \approx (-11^{\circ},-3.5^{\circ})$, due to the much higher differential reddening and disk contamination toward those sightlines. The photometric coverage used in this work is shown in Figure \ref{Fig:ogle3-blg-fields}. Of the 263 fields used, 37 are toward northern latitudes. More detailed descriptions of the instrumentation, photometric reductions and astrometric calibrations are available in  \citet{2003AcA....53..291U}, \citet{2008AcA....58...69U} and \citet{2011AcA....61...83S}. OGLE-III photometry is available for download from the OGLE webpage \footnote{http://ogle.astrouw.edu.pl/}.

We also make use of data from the Two Micron All Sky Survey (2MASS, \citealt{2006AJ....131.1163S}), which we calibrate on the measurements of \citet{2012A&A...543A..13G}, who used data from the VISTA Variables in The Via Lactea survey (VVV, \citealt{2012A&A...537A.107S}). The calibration is discussed in Section \ref{subsec:JMK2MASS}.

\section{Measuring the Red Clump}
\label{sec:RCmeasurements}
The RC is a prominent, well-populated, and localized feature of Galactic bulge CMDs \citep{1988AJ.....96..884T,1994ApJ...429L..73S}, of which we show two examples in Figure \ref{Fig:markovscript17Paper}. The color, color-dispersion, apparent magnitude, magnitude dispersion and normalization of the RC  vary from sightline to sightline, rendering it a sensitive probe of the reddening toward the bulge, its distance, and its underlying structure. We measure these properties across the OGLE-III bulge sky in the following manner.

Each of 2,104 OGLE-III subfields (eight detectors over 263 fields) used in this work was split into 1, 2, 3, 6, 8, 10, 15 or 21 rectangles. Smaller rectangles were used toward regions of the sky where the surface density of stars was higher.  Our average rectangle size is $ \sim 6\arcmin \times 6\arcmin$. The total number of sightlines is 9,744, though for most of this work we only make use of 9,014 of those sightlines that are no nearer than 7$\arcmin$ or 3 half-light radii to a known Galactic globular cluster \footnote{As searched for using A Galactic Globular Cluster Database:  http://gclusters.altervista.org/index.php, which is based on the Harris catalog \citep{1996AJ....112.1487H}.}, that satisfy our photometric completeness criteria of $(V-I)_{RC} \leq 3.30$ and $I_{RC} \leq 17.70$, and that are not flagged as being highly differentially reddened, or otherwise problematic. We publish the best-fit parameter values for the remaining sightlines but do not incorporate them in our analysis.  

The complexity of bulge CMDs requires that we be careful before fitting a luminosity function. The typical limits to the color-magnitude selection box are given by:
\begin{equation}
  -0.30 < (V-I)-(V-I)_{RC}  
\label{EQ:colorlimit}
\end{equation}
\begin{equation}
  -1.5  < I-I_{RC}   < 1.5.
\label{EQ:maglimit}
\end{equation}
The purpose of the color-magnitude selection is to select as many bulge giants as possible without selecting too many foreground stars, which populate a sequence of stars $\sim$0.6 mag bluer than the RC at the brightness of the RC, but that merges with the bulge red giant (RG) branch at a luminosity $\sim$2 mag fainter than the RC \citep{1997ApJ...485..611K}. The color range is therefore adjusted when the foreground main-sequence stars either have very similar or very distinct colors from the bulge RG branch.

\subsection{Mean Magnitude, Magnitude Dispersion and Normalization \\ of the Red Clump}
The standard methodology for investigations of the RC toward various stellar systems is the Paczynski-Stanek equation \citep{1998ApJ...494L.219P}:
\begin{equation}
N(I)dI = a + b(I-I_{RC}) + c(I-I_{RC})^2 + \frac{N_{RC}}{{\sigma}_{RC}\sqrt{2\pi}}
\exp\biggl[ -\frac{1}{2} \frac{(I-I_{RC})^2}{2\sigma_{RC}^2} \biggl],
\end{equation}
where $N(I)dI$ is the number of stars at magnitude $I$ in an interval of length $dI$, the Gaussian parameters $I_{RC}$, $\sigma_{RC}$, and $N_{RC}$ measure the mean magnitude, magnitude dispersion, and number of RC stars, and a quadratic polynomial is fit for the underlying luminosity function of red giant (RG) stars. Though the broad application  of the Paczynski-Stanek equation demonstrates its versatility, we modify it to enhance our accuracy and precision.  

\begin{figure}[H]
\begin{center}
\includegraphics[totalheight=0.6\textheight]{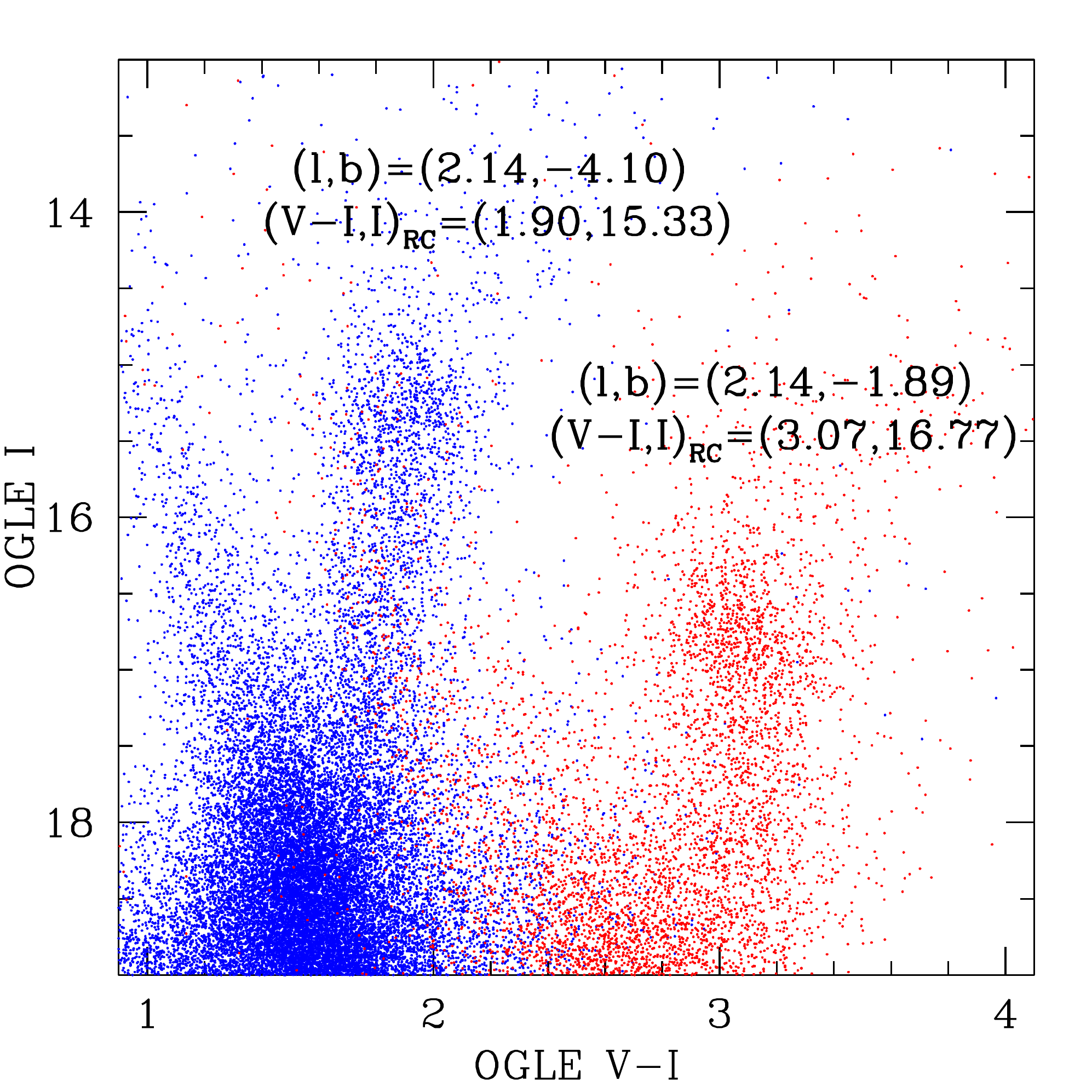}
\end{center}
\caption{OGLE-III CMDs for two distinct sightlines shown on the same figure. The morphology of the CMDs are similar, with both having a foreground disk main-sequence component to their left with a bulge RG branch including a RC to their right. However, the stars toward $(l,b)=(2.14^{\circ},-1.89^{\circ})$ (red), are redder and fainter than the analogous stars toward $(l,b)=(2.14^{\circ},-4.10^{\circ})$ (blue), due to higher interstellar extinction. The RC toward the former sightline is measured to be 1.17 mag redder and 1.44 mag fainter. }  
\label{Fig:markovscript17Paper}
\end{figure}

\begin{figure}[H]
\begin{center}
\includegraphics[totalheight=0.7\textheight]{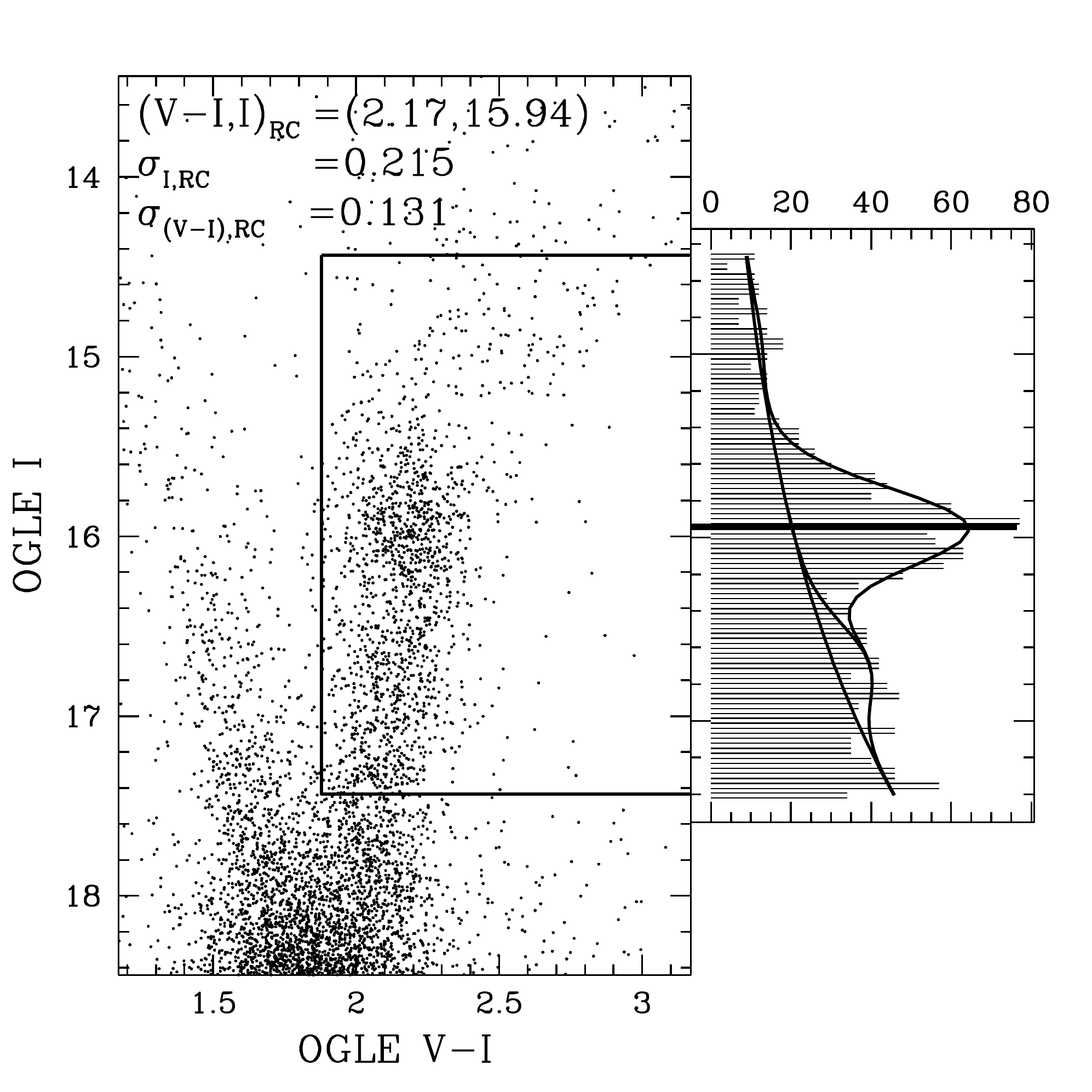}
\end{center}
\caption{OGLE-III CMD toward $(l,b)=(-2.29^{\circ},-3.12^{\circ})$ shown in the left panel. The best-fit values of the color and magnitude of the RC, the magnitude dispersion and the color dispersion are shown on the top-left of the left panel. The color-magnitude selection is denoted by the thick black lines. The magnitude histogram of stars in the color-magnitude selection box is shown to the right of the CMD, on the same scale to the vertical axis, along with a model fit for the RG, RC+RG and total RG+RGBB+RC+AGBB. The parameter values for the RGBB and AGBB is measured in \citet{2011arXiv1109.2118N}} 
\label{Fig:markovscript17Paper2}
\end{figure}

As in some previous works \citep{2010ApJ...721L..28N,2011ApJ...730..118N,2011ApJ...736...94N,2011arXiv1109.2118N,2011arXiv1106.0005N}, we fit the luminosity function of the RG branch to a 2-parameter exponential rather than a 3-parameter quadratic. The reduced number of free parameters makes the fitting routine more stable. We found that fitting a quadratic to the RG branch can lead to catastrophic errors. That is because for large values of $\sigma_{RC}$, the Gaussian becomes degenerate with the quadratic term, leading to even larger values of $\sigma_{RC}$ and $N_{RC}$ at the expense of an unphysical, negative normalization to the RG branch. The exponential satisfies the physically-motivated condition of being both a strictly positive and strictly increasing function of magnitude, which stabilizes it. It is also sound theoretically, as stellar models actually predict an exponential distribution to the magnitudes of RG stars outside the red giant branch bump (RGBB) \citep{1989A&A...216...62C}. We also accounted for the RGBB and asymptotic giant branch bump (AGBB). We parameterized the luminosity function as follows:
\begin{eqnarray}
N(I)dI = A\exp\biggl[B(I-I_{RC})\biggl] + 
 \frac{N_{RC}}{\sqrt{2\pi}\sigma_{RC}}\exp \biggl[{-\frac{(I-I_{RC})^2}{2\sigma_{RC}^2}}\biggl] 
+ \nonumber \\
 \frac{N_{RGBB}}{\sqrt{2\pi}\sigma_{RGBB}}\exp \biggl[{-\frac{(I-I_{RGBB})^2}{2\sigma_{RGBB}^2}}\biggl]  
+ \frac{N_{AGBB}}{\sqrt{2\pi}\sigma_{AGBB}}\exp \biggl[{-\frac{(I-I_{AGBB})^2}{2\sigma_{AGBB}^2}}\biggl]  \nonumber \\
\label{EQ:Exponential},
\end{eqnarray}
 We fixed the parameters of the RGBB and AGBB to their mean values (from \citealt{2011arXiv1109.2118N}) to minimize the systematic effect of extra parameters on the fits:
\begin{eqnarray}
N_{RGBB} = 0.201\times N_{RC} \nonumber \\
N_{AGBB} = 0.028\times N_{RC} \nonumber \\
I_{RGBB} = I_{RC}+0.737 \nonumber \\
I_{AGBB} = I_{RC}-1.07 \nonumber \\
\sigma_{RGBB} = \sigma_{AGBB} = \sigma_{RC}
\label{EQ:bumparameters}
\end{eqnarray}
We imposed the constraint that the integral $\int N(I)dI$ be equal to the number of stars in each fit, and thus we have four free parameters: $I_{RC}$, $\sigma_{RC}$, $B$, and $N_{RC}/A$. We contrast two CMDs in Figure \ref{Fig:markovscript17Paper} specifically selected to demonstrate the effects of reddening and extinction. In Figure \ref{Fig:markovscript17Paper2}, we show the CMD for a typical sightline, as well as our color-magnitude selection box for the fit, the corresponding magnitude histogram and best-fit to the RC+RG+RGBB+AGBB luminosity function in $I$.

\subsection{$(V-I)$ Color and Color-Dispersion of the Red Clump}
The color of the RC is first assumed to be that of the nearest of 92,000 values of $(V-I)_{RC}$ measured for the study of \citet{2010ApJ...721L..28N}. The measurement of $(V-I)_{RC}$ is then slightly adjusted in this work. We start with the old value, and then measure the position of the RC in $I$ by fitting Equation \ref{EQ:Exponential}. Two colors are subsequently fit for, one for the RC, and one for the position of the foreground disk component at the brightness of the RC (typically $\sim$0.6 mag bluer than the RC), easily discernible in Figures  \ref{Fig:markovscript17Paper} and \ref{Fig:markovscript17Paper2}. Every star in the CMD is then assigned to the closer of these two colors, and the two colors selected are those that minimize the weighted variance of the difference between the colors of stars and the nearest of the two trial colors (RC and foreground disk component), where the weights $W_{i}$ are given by:
\begin{equation}
W_{i} = \frac{\frac{N_{RC}}{\sqrt{2\pi}\sigma_{RC}}\exp \biggl[{-\frac{(I-I_{RC})^2}{2\sigma_{RC}^2}}\biggl]}{A\exp\biggl[B(I-I_{RC})\biggl] +  \frac{N_{RC}}{\sqrt{2\pi}\sigma_{RC}}\exp \biggl[{-\frac{(I-I_{RC})^2}{2\sigma_{RC}^2}}\biggl]  },
\end{equation}
and thus in practice only stars satisfying $|I - I_{RC}| \lesssim \sigma_{RC}$ contribute to the fit to the color and color-dispersion. We recursively removed 2.5$\sigma$ outliers from the fit to $(V-I)_{RC}$. The values of the color $(V-I)_{RC}$ and color-dispersion $\sigma_{(V-I)_{RC}}$ are adopted once no outliers remain. If the resulting color disagrees with the color first assumed by 0.03 mag or more, we redo the fit to the magnitude distribution and subsequently to the color with an updated color-magnitude selection box, so that the number of stars bluer than the RC included in the fit is independent of any potentially incorrect initial guess of the RC colour. If the fit fails to converge after 5 iterations, the 0.03 mag condition is relaxed, and the sightline is flagged as problematic and not incorporated within any of our analysis.

\subsection{$(J-K_{s})_{RC}$ of the Red Clump}
\label{subsec:JMK2MASS}
\citet{2012A&A...543A..13G} measured $(J-K_{s})_{RC}$ across the viewing area $(-10 \leq l \leq +10.2, -10 \leq  b \leq +5)$ using data from the VVV survey \citep{2012A&A...537A.107S}. These results are used in our work, and cover most of the OGLE-III bulge viewing area. 

For the remaining part of the sky, we used $(J-K_{s})_{RC}$ measurements from 2MASS data \citep{2006AJ....131.1163S} that we calibrated on the measurements of \citet{2012A&A...543A..13G}.  The magnitude limit of 2MASS is typically brighter than the bulge RC. However, 2MASS does reach the RG stars that are at the color of the RC and brighter than the RC. We cross-matched the OGLE-III and 2MASS source catalogs, yielding a $VIJK_{s}$ photometry list for all RG stars brighter than the RC. For RG stars satisfying $|(V-I)_{RG} - (V-I)_{RC}| \leq 0.33$, we computed the linear relations $(I-J),(I-K_{s}) = a + b{\times}((V-I)_{RG} - (V-I)_{RC})$. The difference of the two relations then yields $(J-K_{s})$ for RG stars at the $(V-I)$ color of the RC. This methodology was first developed as a means to study gravitational microlensing events toward the bulge (\citealt{2010ApJ...713..837B,2009ApJ...698L.147G}). 

\begin{figure}[H]
\begin{center}
\includegraphics[totalheight=0.5\textheight]{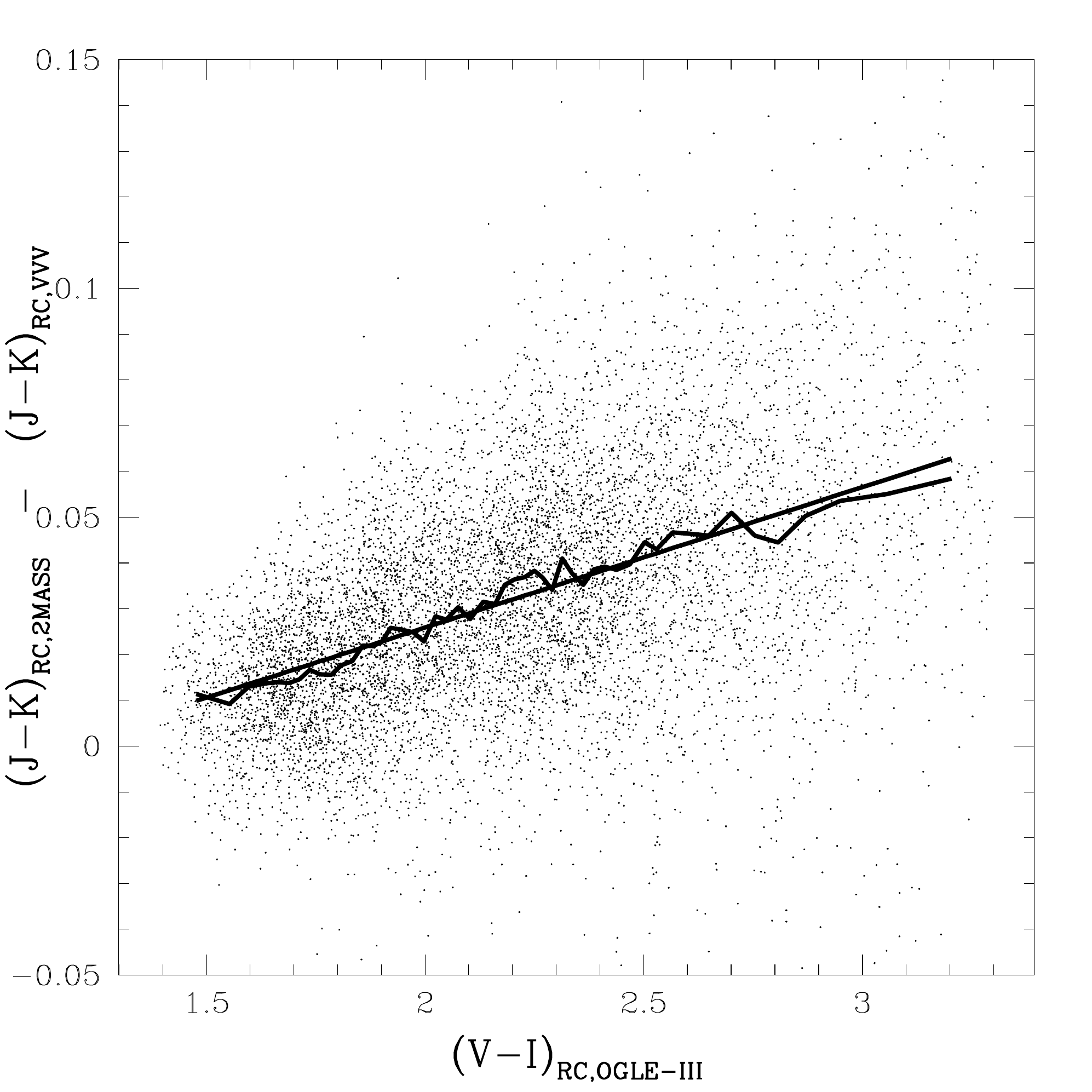}
\end{center}
\caption{Difference in $(J-K_{s})_{RC}$ inferred from 2MASS photometry with that directly measured from deep VVV photometry by \citet{2012A&A...543A..13G}. The relationship is $(J-K_{s})_{\rm{RC,2MASS}} - (J-K_{s})_{\rm{RC,VVV}} = -0.035 + 0.031{\times}(V-I)_{\rm{RC,OGLE-III}}$, which we subsequently correct for. Points denote each measurement. Both the medians to 50 bins and the best-fit line are shown by thick black curves.} 
\label{Fig:ColorError}
\end{figure}

\section{Calibration of the Intrinsic Luminosity Parameters of the Red Clump, $(V-I)_{RC,0}$, $M_{I,RC}$, $\sigma_{I,RC,0}$ and $(J-K_{s})_{RC,0}$}
\label{sec:Calibration}
We adopt $(M_{I,RC}, \sigma_{I,RC,0}, (V-I)_{RC,0}, \sigma_{(V-I)_{RC}}, (J-K_{s})_{RC,0}) = (-0.12, 0.09, 1.06, 0.121, \newline 0.66)$ for the mean absolute magnitude and magnitude dispersion in $I$, intrinsic $(V-I)$ color, intrinsic $(V-I)$ color-dispersion, and intrinsic $(J-K_{s})$ color of the RC.  Our justifications for $M_{I,RC}$, $(V-I)_{RC,0}$, and $(J-K_{s})_{RC,0}$ is provided below. The intrinsic color dispersion of the RC is measured to be $\sigma_{(V-I)_{RC,0}} = 0.121$ in Section \ref{sec:ReddeningLawEstimates}, and an estimate of the intrinsic magnitude dispersion ${\sigma}_{I,RC,0} \approx 0.14-0.17$ was provided by \citet{2011arXiv1106.0005N}, but is revised downward to ${\sigma}_{I,RC,0} = 0.09$ mag in this work. 

\subsection{$(V-I)_{RC,0}$ From the Spectroscopy of Microlensed Dwarf and Subgiant Stars}
Spectroscopic investigations of high-magnification microlensing events toward the bulge have yielded detailed abundances and reddening-independent temperatures for bulge main-sequence turnoff and subgiant stars \citep{2007ApJ...655L..33J,2008ApJ...685..508J,2008ApJ...682.1029C,2009ApJ...699...66C,2010ApJ...709..447E,2009A&A...499..737B,2010A&A...512A..41B,2011A&A...533A.134B}. The intrinsic $(V-I)$ colors of these stars can be estimated from the measured temperatures and metallicities and the empirical [Fe/H]-color-$T_{\rm{eff}}$ calibration of \citet{2010A&A...512A..54C}. This intrinsic color can be compared to the observed color to obtain the reddening. The assumption that the reddening to the source equals the reddening to the RC centroid yields $(V-I)_{RC,0} = 1.06$ \citep{2011A&A...533A.134B}.

\subsection{$(V-I)_{RC,0}$ and $M_{I,RC}$ by Linear Interpolation with 47 Tuc and NGC 6791}
\label{sec:Tuc6791}
We measure the properties of the RC for two old stellar populations that are well-calibrated to provide an empirical relation.

We first use photometry for the Galactic globular cluster 47 Tuc (NGC 104) from the catalog of \citet{2007AJ....133.1658S}. The $VI$ magnitudes are obtained by converting from the original $F606W$ and $F814W$ photometry taken with the \textit{HST}. The limits used to identify RC stars are the same as in \citet{2011ApJ...736...94N}: $(0.82 \leq V-I \leq 0.98,\,13.80 \leq V \leq 14.20)$, thus delineating a box on the CMD containing 546 stars. We measure $( (V-I)_{RC},\,I_{RC},\,\sigma_{I,RC})_{\rm{47\,\,Tuc}} = (0.913,\,13.090,\,0.062)$.  \citet{2010AJ....139..329T} precisely measured the properties of an eclipsing binary (EB) pair in the cluster, and determined $(m-M)_{V}=13.35 \pm 0.08$. Assuming a cluster reddening $E(B-V)=0.04$ \citep{1996AJ....112.1487H} and a standard reddening law \citep{1989ApJ...345..245C,1994ApJ...422..158O}, we find $( (V-I)_{RC,0},\,M_{I,RC},\,\sigma_{I,RC,0})_{\rm{47\,\,Tuc}} = (0.861,\, -0.208,\,0.062)$. For the cluster metallicity, we adopt [Fe/H]$=-0.76$ \citep{2008AJ....135.1551K}.

For the open cluster NGC 6791, we use the photometric catalog of \citet{2003PASP..115..413S}, which was updated for the work of \citet{2012arXiv1205.4071B} and now includes corrections for differential reddening. We draw a box around the RC using the limits $(1.25 \leq V-I \leq 1.40,\,13.05 \leq I \leq 13.35)$, thus delineating a box on the CMD containing 22 stars. We measure $( (V-I)_{RC},\,I_{RC},\,\sigma_{I,RC})_{\rm{NGC\,\,6791}} = (1.326,\,13.243,\,0.050)$. \citet{2011A&A...525A...2B,2012arXiv1205.4071B} precisely measured the properties of 3 eclipsing binary pairs in the cluster as well as the cluster CMD, and found $E(V-I) = 0.174$, $(m-M)_{V}=13.51\pm0.06$, and [Fe/H]$=+0.29$. We thus infer $( (V-I)_{RC,0},\,M_{I,RC},\,\sigma_{I,RC,0})_{\rm{NGC\,\,6791}} = (1.152,\, -0.093,\,0.050)$. 

The assumption of linear population effects in the metallicity range bracketed by 47 Tuc and NGC 6791 yields the following empirical relations:
\begin{eqnarray}
(V-I)_{RC,0} = 1.09 + 0.277\times(\rm{[Fe/H]}-0.05) 
\label{EQ:IRCempirical} \\
M_{I,RC} = -0.12 + 0.110\times(\rm{[Fe/H]}-0.05)
\label{EQ:VMIRCempirical}
\end{eqnarray}

The bulge RC has a measured metallicity distribution function with mean $\overline{\rm{[Fe/H]}}=+0.05$ \citep{2011A&A...534A..80H}. The resulting RC parameters are $((V-I)_{RC,0},\,M_{I,RC})_{\rm{Bulge}} = (1.09,\, -0.12)$. This derivation implicitly assumes that other parameters that effect HB morphology, such as age, $\alpha$-enhancement, helium, and binarity, also behave in a linear or nearly-linear manner in the metallicity interval bracketed by 47 Tuc and NGC 6791. The potential impact of these uncertainties is discussed in Section \ref{sec:TheoreticalUncertainties}.

\subsection{$(V-I)_{RC,0}$, $M_{I,RC}$ from Stellar Models, a \textit{Hipparcos} calibration, and the Bulge RC Spectroscopic Metallicity Distribution Function}
\citet{2001MNRAS.323..109G} and \citet{2002MNRAS.337..332S} constructed a grid of horizontal branch stellar models to predict the functional dependence of the properties of the RC on various stellar population parameters. For the Galactic bulge, they combined the mean $\alpha$-enhancement and metallicity of \citet{1994ApJS...91..749M} ([Fe/H]$=-0.22$, [$\alpha$/Fe]$=+0.35$) and assumed an age range of 8 to 12 Gyr to predict that the bulge RC should be 0.06 mag redder in $(V-I)$ and 0.01 mag fainter in $I$ than the stars of the solar neighborhood, which could be calibrated by using distances from the \textit{Hipparcos} catalog \citep{1997A&A...323L..49P}.

We update this estimate by using the same grid of stellar model outputs,  and by assuming the same age range. We use the metallicity distribution function of \citet{2011A&A...534A..80H}, who measured 219 spectroscopic abundances in a sample of RC+RG stars with mean $(V-I,I)$ corresponding to the measured position of the RC. They obtained a mean metallicity of [Fe/H]$=+0.05$, and mean [Mg/Fe]$=+0.15$, which we use as a proxy for the mean $\alpha$-abundance. The 0.27 dex higher [Fe/H] and approximately half-as-high [$\alpha$/Fe] shifts the bulge corrections, to 0.20 mag redder in $(V-I)$ and 0.10 mag fainter in $I$ relative to that of the solar neighborhood. For the \textit{Hipparcos} stars, the value of $(V-I)_{RC,0} \approx 1.01$ \citep{1998ApJ...494L.219P}, and $M_{I,RC} = -0.22 \pm 0.03$  \citep{2008A&A...488..935G}, yielding $((V-I)_{RC,0},M_{I,RC}) = (1.21,\,-0.12)$. It is not clear why this result from theory is substantially redder than our empirical estimates. It could be due to a difference in the mean metallicity of sightlines studied by \citet{2011A&A...533A.134B} and that studied by \citet{2011A&A...534A..80H}. Another possibility is a different mapping between [Fe/H], the different $\alpha$ elements, age, and helium for bulge stars than that assumed by \citet{2001MNRAS.323..109G}.

\subsection{$\sigma_{I,RC,0}$ from Stellar Models and the Spectroscopic Metallicity Distribution Function}
The intrinsic magnitude dispersion of the RC, $\sigma_{I,RC,0}$, is the magnitude dispersion that the bulge RC would have if the Galactic bulge were geometrically thin and if there was no differential extinction, i.e.:
\begin{equation}
\sigma_{I,RC,0}^2 = \sigma_{I,RC}^2 - \sigma_{\mu}^{2} - \sigma_{A_{I}}^2, 
\end{equation}
where $\sigma_{\mu}$ is the dispersion in distance modulus, and $\sigma_{A_{I}}$ is the dispersion in extinction. Understanding of $\sigma_{I,RC,0}$ is essential if one is to constrain $ \sigma_{\mu}$, a fundamental probe of Galactic structure \citep{1997ApJ...477..163S,2008A&A...491..781C}. \citet{2011arXiv1106.0005N} estimated $\sigma_{I,RC,0} \approx 0.17$, but that was done before realizing one of the findings of this work: that differential extinction can exceed values of 0.10 mag on scales as small as $\sim 6 \arcmin \times 6 \arcmin$ (see Section \ref{subsec:differentialreddening}). We list three estimates: \begin{itemize}
\item \citet{2001MNRAS.323..109G} used a large, detailed grid of stellar tracks, the metallicity distribution function of \citet{1994ApJS...91..749M}, and a constant, metallicity-independent star-formation rate over the range $8 \leq (t/\rm{Gyr}) \leq 12$ to estimate $\sigma_{I,RC,0} = 0.107$ mag.
\item We use the synthetic HB calculator of the BaSTI stellar database      \citep{2004ApJ...612..168P,2006ApJ...642..797P}  to estimate  $\sigma_{I,RC,0} = 0.031$ for a simple, old, metal-rich stellar population. We quadratically add the effect of the dispersion in metallicity of the bulge (0.40 dex, \citealt{2011A&A...534A..80H}) and the predicted evolutionary effect from the BaSTI database of variable metallicity on the RC luminosity, $dM_{I}/d\rm{[Fe/H]}=0.20$ mag dex$^{-1}$, to predict $\sigma_{I,RC,0} = 0.086$ mag.     
\item From Section \ref{sec:Tuc6791}, we have $\sigma_{I,RC,0} = 0.055$ mag for a simple, old, metal-rich stellar population, and $dM_{I}/d\rm{[Fe/H]}=0.11$ mag dex$^{-1}$. This yields an estimate of  $\sigma_{I,RC,0} = 0.071$ mag.
\end{itemize}
The mean of the three estimates, $\sigma_{I,RC,0} = 0.09$ mag, is assumed in this work. 

The estimates of  $\sigma_{I,RC,0}$ derived here do not account for any possible presence of a secondary RC due to stars with non-generate cores \cite{1999MNRAS.308..818G}, as we do not expect such a population to be substantial in the bulge. If present, however, the brightness dispersion would be increased.

\subsection{$(J-K_{s})_{RC,0}=0.66$ from Red Giant Color-Color Relations and Relative Reddening Expectations}

An estimate of $(J-K_{s})_{RC,0}$ can be derived by combining the estimate $(V-I)_{RC,0} = 1.06$ with the numerous, precision measurements of RG color-color relations of \citet{1988PASP..100.1134B}. We first use the approximation that:
\begin{equation}
K_{s} = K + 0.1{\times}(H-K).
\end{equation}
we then interpolate rows 5 and 6 of their Table III to derive that:
\begin{equation}
(J-K_{s})_{0} = 0.620 + 0.625{\times}((V-I)_{0} - 1.00),
\end{equation}
yielding $(J-K_{s})_{RC,0} = 0.66$.

A second estimate can be derived by studying the distribution of inferred $E(J-K_{s})/E(V-I)$ values (derived in Section \ref{sec:RJKVI}) as the assumed values for $(J-K_{s})_{RC,0}$ and $(V-I)_{RC,0}$ are varied systematically.  For small errors in the intrinsic colors, denoted ${\Delta}(J-K_{s})_{0} = (J-K_{s})_{0,\rm{Inferred}} - (J-K_{s})_{0,\rm{True}}$ and ${\Delta}(V-I)_{0} = (V-I)_{0,\rm{Inferred}} - (V-I)_{0,\rm{True}}$:
\begin{equation}
\frac{E(J-K_{s})}{E(V-I)}_{\rm{Inferred}} = \frac{E(J-K_{s})}{E(V-I)}_{\rm{True}}{\times}\biggl(1 + \frac{{\Delta}(V-I)_{0}}{E(V-I)} - \frac{{\Delta}(J-K_{s})_{0}}{E(J-K_{s})} \biggl),
\label{EQ:colorerrors}
\end{equation}
from which it follows that incorrect assumptions as to the values of $(J-K_{s})_{RC,0}$ and $(V-I)_{RC,0}$ will cause the inferred values of $E(J-K_{s})/E(V-I)$ to be correlated with the reddening. We do a grid search and find that values of the intrinsic colors that remove the correlation  between $E(V-I)$ and $E(J-K_{s})/E(V-I)$, i.e., $cor(E(V-I),E(J-K_{s})/E(V-I))=0$, satisfy the following relation:
\begin{equation}
(J-K_{s})_{0} =  0.641 + 0.340{\times}((V-I)_{0} - 1.00).
\label{EQ:covarianceRJKVI}
\end{equation}
For $(V-I)_{RC,0} = 1.06$, Equation (\ref{EQ:covarianceRJKVI}) predicts $(J-K_{s})_{RC,0}=0.66$. Alternatively, requiring $cor(E(V-I),E(J-K_{s})/E(V-I)) = \pm 0.1$ would shift $(J-K_{s})_{RC,0}$ by 0.01, or $(V-I)_{RC,0}$ by 0.03, where a lower value of $(J-K_{s})_{RC,0}$ or a higher value of $(V-I)_{RC,0}$ would necessitate a lower value of $E(J-K_{s})/E(V-I)$ at higher $E(V-I)$. We show later in the text that larger absolute values of the correlation are unphysical, and thus this method should determine $(J-K_{s})_{RC,0}=0.66$ to an accuracy of 0.01, for a given value of $(V-I)_{RC,0}$.


\subsection{Estimating the RC Intrinsic Luminosity in $H$ and $K_{s}$ from our Calibrations}
\label{sec:IRestimates}
We include estimates of the RC luminosity in $H$ and $K$ even though these are not directly relevant to our study, as these may be useful elsewhere, for example in microlensing studies that combine information from several bandpasses and would benefit from an internally consistent set of calibrations.

We combine our calibrations of $M_{I}$ and $V-I$ with the empirical RG color-color relations to estimate $(V-K_{s})_{RC}=2.44$, $(H-K_{s})_{RC}=0.09$, $M_{H,RC}=-1.41$ and $M_{K_{s},RC}=-1.50$.
	
\subsection{Theoretical Estimates of the Effect of Population Uncertainties \\ on Red Clump Parameters}
\label{sec:TheoreticalUncertainties}
It is worthwhile to ask how the derived luminosity parameters of the bulge RC would vary if our  population assumptions were to change. For this task we use the BaSTI stellar database \citep{2004ApJ...612..168P,2006ApJ...642..797P}, whose breadth allows us to inspect the effect of several stellar parameters. The predicted effects of changing metallicity, age, and variable mass loss are computed using the synthetic HB generator, which interpolates between canonical tracks. We note that helium-enrichment is coupled to [Fe/H] as their models assume ${\Delta}Y/{\Delta}Z = 1.4$. For helium-enrichment beyond (or below) this quantity, we must use the much coarser grid of He-enhanced models. We read off the difference in mean position of the HB tracks for the [Fe/H]$=-0.70$, t$=10$ Gyr, Y$= 0.30$ and the  [Fe/H]$=-$0.70, t$=$10 Gyr, Y$=$0.35 models. 

We thus obtain the following predicted differential relations for the brightness and color of the RC:
\begin{equation}
M_{I,RC} \propto +0.20\rm{[Fe/H]} + 0.015(t/\rm{Gyr}) - 0.029(100Y-100Y_{ss}) + 0.06({\delta}{\Delta}M/0.1M_{\odot}),
\label{EQ:Iterms}
\end{equation}
\begin{equation}
(V-I)_{RC,0} \propto 0.27\rm{[Fe/H]} - 0.013(100Y-100Y_{ss} ),
\label{EQ:colorterms}
\end{equation}
where $Y_{ss}$ is the scaled-solar helium abundance at metallicity [Fe/H],  ${\delta}{\Delta}M$ parameterizes any potential difference in the mass loss relative to that assumed by the BaSTI database. Some readers may wonder why Equation \ref{EQ:colorterms} has fewer terms than Equation \ref{EQ:Iterms}: for sufficiently high-metallicity, the mean color of the RC star becomes nearly independent of small changes in the total mass of the star, and thus either the mass-loss or the age of the progenitor.

These predicted variations in the luminosity parameters of the RC can lead to errors in the distance determinations. We derive this below. Due to the proportionality between reddening and extinction, errors in the color will lead to errors in the dereddened apparent magnitude and thus distance modulus:
\begin{equation}
{\mu} = I_{RC} - M_{I,RC} - R_{I}{\times}[ (V-I)_{RC} - (V-I)_{RC,0}],
\label{EQ:colorterms2}
\end{equation}
setting $R_{I}=1.22$, an average derived later in this work, yields:
\begin{equation}
{\delta}{\mu} \propto 0.13{\delta}\rm{[Fe/H]} + 0.013(100Y-100Y_{ss})-0.015({\delta}t/\rm{Gyr})-0.06({\delta}{\Delta}M/0.1M_{\odot}).
\label{EQ:colorterms3}
\end{equation}
An unaccounted increase of either 0.02 in $(Y-Y_{ss})$ or 0.2 dex in [Fe/H] will thus yield an decrease in the inferred distance of $\sim$100 pc, i.e. sight lines with higher metallicity or relatively enhanced-helium will appear closer than they really are.

\section{The Reddening Law in $V$, $I$, $J$ and $K_{s}$: Theoretical Expectations}
\label{Sec:ReddeningTheory}
\citet{1989ApJ...345..245C} combined data from several sources and found that the extinction law over the wavelength range $3.5\,{\mu}m \geq \lambda \geq 0.125\,{\mu}m$ could be parameterized by a single variable, $R_{V} =A_{V}/E(B-V)$. Thus, if one knows $R_{V}$ and a reddening value toward a particular sightline, one can derive the extinction for each wavelength in that calibrated range. The equations of \citet{1989ApJ...345..245C} were updated by \citet{1994ApJ...422..158O}, who obtained additional data. 

We convolve $V$ and $I$ filters of the Landolt photometric system \citep{1992AJ....104..372L} that OGLE-III photometry is calibrated on \citep{2011AcA....61...83S} with a 4700 K blackbody curve, typical of bulge RC stars \citep{2011A&A...534A..80H}. We obtain effective wavelengths of $0.546\,{\mu}$m and $0.804\,{\mu}$m for the $V$ and $I$ filters. For the $J$ and $K_{s}$ filters, we do as \citet{2012A&A...543A..13G} and \citet{2005ApJ...619..931I} , and respectively adopt $1.240\,{\mu}$m and $2.164\,{\mu}$m. These effective wavelengths are a little different than those adopted by \citet{1998ApJ...500..525S}, as those authors assumed the spectral energy distribution of an average elliptical galaxy for the source. 

The standard value of $R_{V}$ for the interstellar medium is $R_{V} = 3.1$ \citep{1975A&A....43..133S,1978ApJ...223..168S}. With the parameterization of \citet{1989ApJ...345..245C}, the predicted reddening terms are $R_{I}=A_{I}/E(V-I)=1.424$ and $R_{JKVI}=E(J-K_{s})/E(V-I)=0.407$, whereas they are $R_{I}=1.481$ and $R_{JKVI}=0.416$ with the parameterization of \citet{1994ApJ...422..158O}. Given these numbers, we adopt $R_{I}=1.45$ and $R_{JKVI}=0.41$ as the standard values to which we compare our results. 

We comment on two additional uncertainties. The first is that due to instrumental uncertainty in the effective central wavelengths of the filters: a change of 1 nm in the effective wavelengths, approximately equivalent to a temperature change of 750 K,  would modify the value of $R_{I}$ by $\sim$0.01 and $R_{JKVI}$ by $\sim$0.004. The second is that the effective wavelengths of the filters are modified when convolved with a foreground extinction. To gauge this effect, we assume $R_{V}=3.1$, the equations of \citet{1994ApJ...422..158O}, and apply 5 magnitudes of extinction in $V$. We find that $R_{I}$ drops by $\sim$0.02 and $R_{JKVI}$ by $\sim$0.007. These effects are real, but not large.

\section{Reddening Derived from the Mean Color of the Red Clump}
\label{sec:Reddening}
The reddening can be derived by taking the difference between the observed and intrinsic value of $(V-I)_{RC}$:
\begin{equation}
E(V-I) = (V-I)_{RC} -  (V-I)_{RC,0} = (V-I)_{RC} - 1.06.
\label{EQ:ReddeningDefinition}
\end{equation}
 The distribution of reddening for  9,014 sightlines is shown as a color-coded map in the top panel of Figure \ref{Fig:Mapper17D4Paper4Maps}.

\subsection{The Scale of Differential Reddening}
\label{subsec:differentialreddening}
Differential reddening, though less investigated than mean reddening, is of great importance to Galactic bulge studies. Spectroscopic investigations of red giants assume photometric temperatures and gravities \citep{2006ApJ...636..821F,2008A&A...486..177Z,2011ApJ...732L..36D,2011A&A...534A..80H,2011ApJ...732..108J}, and thus knowledge of the scale of differential reddening is a prerequisite to making more detailed analyses of the bulge giant metallicity distribution function. Microlensing investigations also rely on colors to constrain the properties of planet-hosting stars \citep{1999ApJ...512..672A,2004ApJ...603..139Y}.

\begin{figure}
\begin{center}
\includegraphics[totalheight=0.8\textheight]{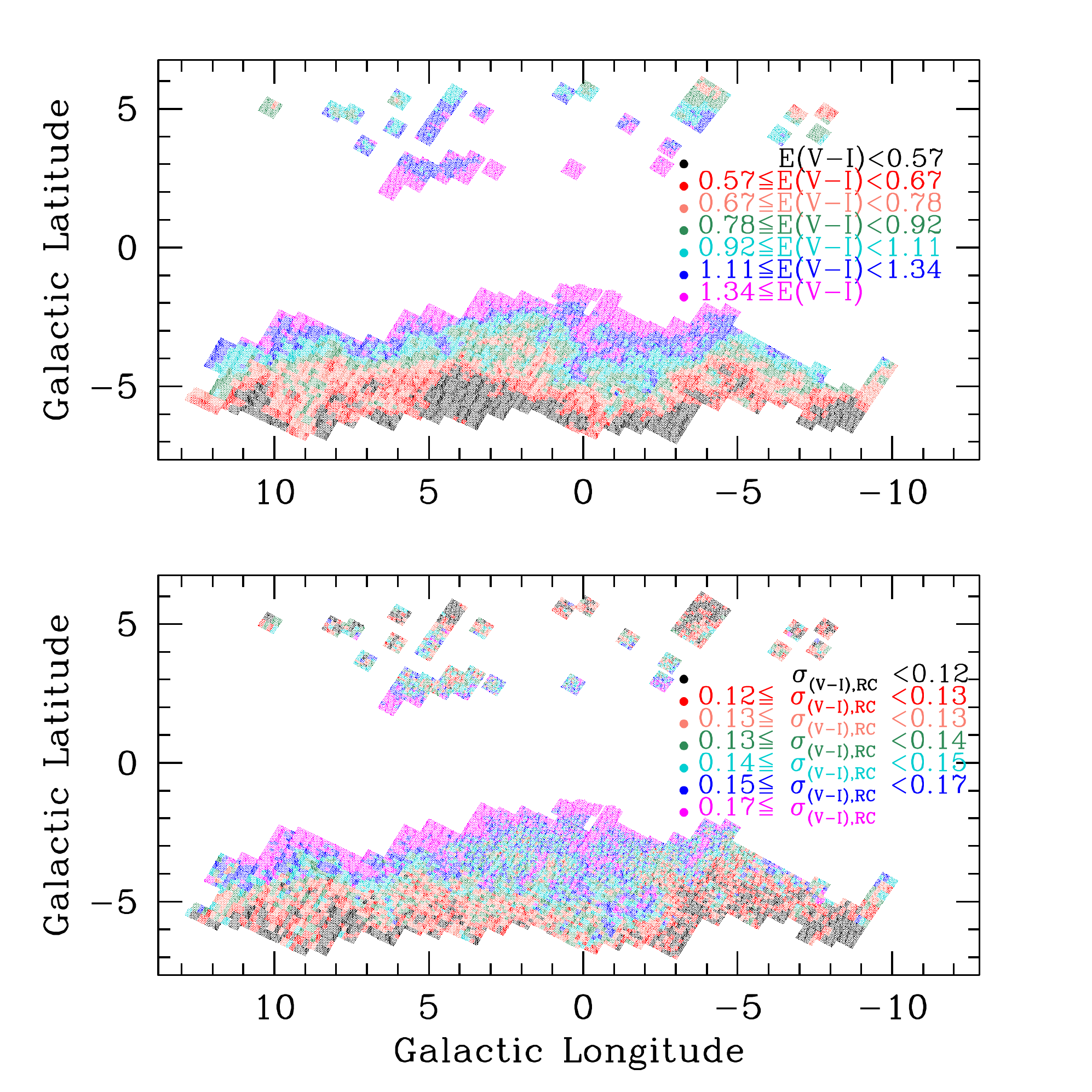}
\end{center}
\caption{TOP: Color-coded reddening map of the Galactic bulge as observed by the OGLE-III $VI$ photometric survey. BOTTOM: Distribution of ${\sigma}_{(V-I),RC}$ as a function of direction. Color-coded map of the color dispersion, $\sigma_{(V-I),RC}$, which is the quadratic sum of the intrinsic color dispersion and the differential reddening.  The distribution of $\sigma_{(V-I),RC}$ looks remarkably similar to the distribution of $E(V-I)$ shown in the top panel, suggesting a functional dependence of differential reddening on total reddening. For both panels, each color codes approximately equal areas. } 
\label{Fig:Mapper17D4Paper4Maps}
\end{figure}

For each of our fields, we have estimated the color dispersion of the RG+RC at the luminosity of the RC, ${\sigma}_{(V-I),RC}$. This quantity will be equal to a quadratic sum of an intrinsic color dispersion and a reddening dispersion:
\begin{equation}
{\sigma}_{(V-I),RC}^2 = {\sigma}_{(V-I),RC,0}^2 + \sigma_{E(V-I)}^2.
\end{equation}
We plot the distribution of ${\sigma}_{(V-I),RC}$ in the bottom panel of Figure \ref{Fig:Mapper17D4Paper4Maps}.

The distribution of color-dispersions looks remarkably similar to the distribution of $E(V-I)$ shown in the top panel of Figure \ref{Fig:Mapper17D4Paper4Maps}. We thus assume a simple parametric form for ${\sigma}_{(V-I),RC}^2$:
\begin{equation}
{\sigma}_{(V-I),RC}^2 = {\sigma}_{(V-I),RC,0}^2 + \biggl[C_{DR1}^2 + C_{DR2}^2\frac{\Delta \Omega}{0.01 \rm{\,deg}^2} \biggl]{\times}E(V-I)^2,
\label{EQ:diffred}
\end{equation}
where $\Delta \Omega$ is the solid angle subtended by the sightline,  $C_{DR1}$ is the proportionality constant between total and differential reddening that is independent of the angular size of the sightline, and $C_{DR2}$ parameterizes the component of differential reddening that depends on the solid angle of the field. We do a three-parameter optimization and find ${\sigma}_{(V-I),RC,0} = 0.121$, $C_{DR1} = 0.070$, and $C_{DR2} = 0.017$, where the fitting was restricted to the 94\% of sightlines with ${\sigma}_{(V-I),RC} \leq 0.25$. As the mean solid angle for our sightlines is 0.011 deg$^{2}$, our results demonstrate that significant differential reddening occurs on scales much smaller than a few arcminutes -- it is essentially granular when reddening is measured with the RC method. The sum of $C_{DR1}$ and $C_{DR2}$ shows that to first order differential reddening averages $\sim$9\% of total reddening, even over the small sightlines used by our investigation.



\subsection{Distribution of Dust Toward the Inner Galaxy}
We estimate the distribution of dust toward the Galactic bulge assuming a simple 2-parameter model: the mean density of dust along the plane $\rho_{D}$ and the scale height $H_{D}$. Thus, the prediction for the reddening toward a given direction is:
\begin{equation}
E(V-I)(l,b) = \int_{0}^{R_{\rm{final}}(l,b)} \rho_{D}\exp\biggl[-r\sin(|b|)/H_{D} \biggl] dr,
\end{equation}
where:
\begin{equation}
R_{\rm{final}}(l,b) = \frac{R_{0}\,\sin({\alpha}) } {\cos(b) \sin(l+\alpha) },.
\end{equation}
We assume a distance to the Galactic center of $R_{0}= 8.4$ kpc \citep{2008ApJ...689.1044G}  and an angle between the bulge's major axis and the Sun-GC line of sight of $\alpha=25^{\circ}$ \citep{2007MNRAS.378.1064R}, both of which are consistent with values estimated later in this work. 

We minimize the sum of the squares of the differences between the predicted and measured reddening. This approximation yields $\rho_{D}=0.427\,\,\rm{mag\,kpc}^{-1}$ and $H_{D}=164$ pc. The resulting scatter is 0.25 mag. This value of the scale height is larger than the 125 pc reported by \citet{2006A&A...453..635M}. This is likely due to the fact a two-parameter model can only go so far. For example, $\rho_{D}$ may be a function of Galactocentric radius.

\begin{figure}
\begin{center}
\includegraphics[totalheight=0.8\textheight]{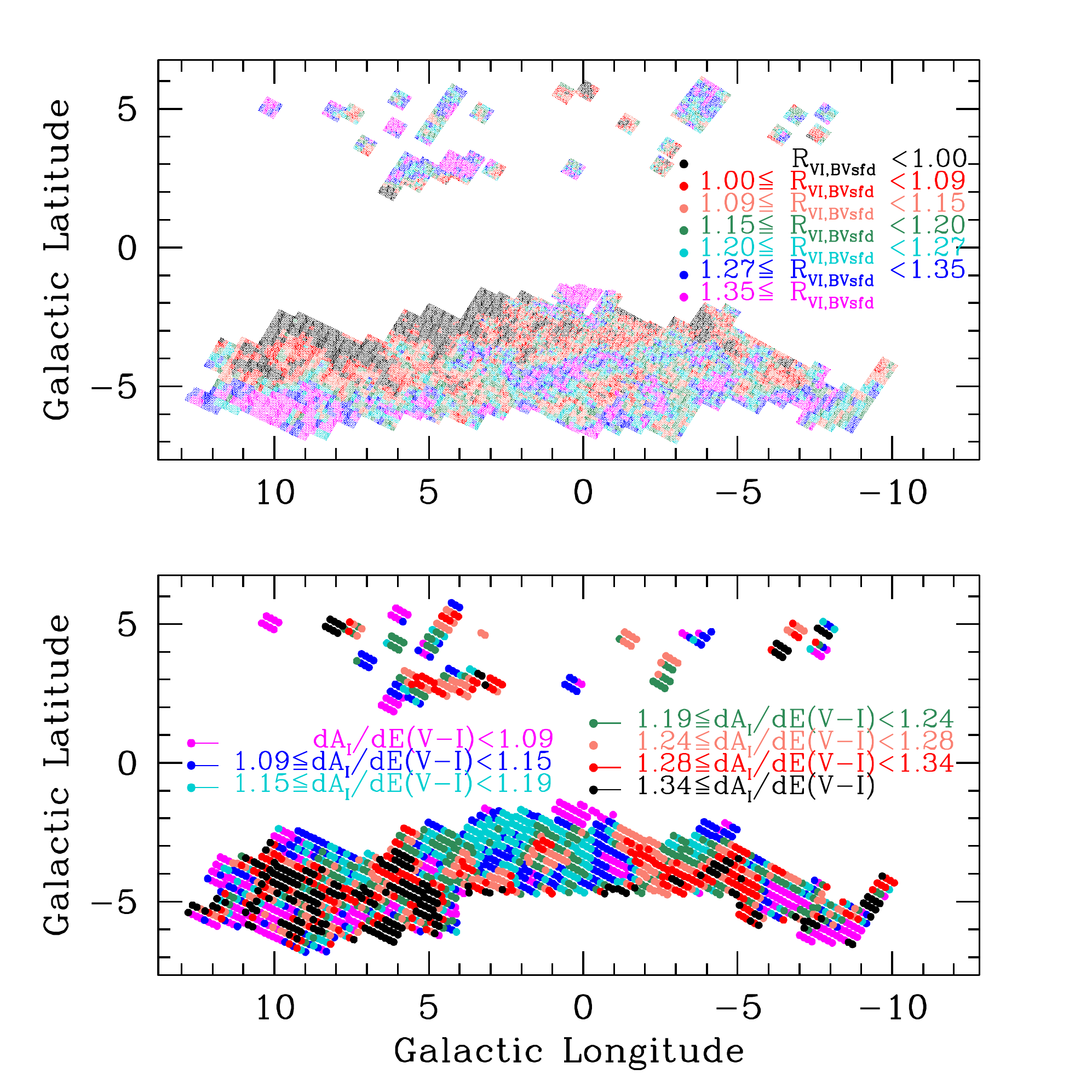}
\end{center}
\caption{TOP: Distribution of $R_{VI,BV\rm{sfd}} = E(V-I)/E(B-V)_{\rm{sfd}}$ as a function of direction. The mean value of $R_{VI,BV\rm{sfd}}=1.179$ is consistent with a mean extinction curve of $R_{V}=2.47$, but the 15\% scatter does not correlate with any known reddening index.  BOTTOM: Color-plot of distribution of measured $dA_{I}/dE(V-I)$ values as a function of position. Each color represents an equal surface area on the sky. More standard values of the reddening law, $dA_{I}/dE(V-I) \geq 1.34$, are observed further from the minor axis. Sightlines with a prominent double-RC have been removed from this analysis.} 
\label{Fig:Mapper17D4Paper4Maps2}
\end{figure}

\subsection{Comparison to the All-Sky Reddening Map of \citet{1998ApJ...500..525S} }
\citet{1998ApJ...500..525S} produced a full-sky reddening map that is now one of the most widely used tools in astronomy. It is thus an important benchmark for comparisons. 

For each of our $E(V-I)$measurements we took the ratio to the nearest 4 pixel interpolation of $E(B-V)$ from \citet{1998ApJ...500..525S}, which we denote $R_{VI,BV\rm{sfd}}$. We find a mean value of $R_{VI,BV\rm{sfd}} = 1.179$ with a standard deviation of 0.173, the distribution is shown as a color-coded map in the top panel of Figure \ref{Fig:Mapper17D4Paper4Maps2}. This is $\sim$17\% lower than the value of $E(V-I)/E(B-V)=1.38$ that \citet{1998ApJ...500..525S} assume for the entire sky. 

The \citet{1998ApJ...500..525S} maps are well-established to trace the extinction poorly in regions with either
high extinction or steep extinction spatial gradients \citet{2012ApJS..201...35N}, and the bulge has plenty of both.  However, there may be some fortuitous cancellation here: the errors in the \citep{1998ApJ...500..525S} values of $E(B-V)$ nearly cancel out with their assumption that $R_{V}=3.1$ toward the bulge, yielding values of $E(V-I)$ (but not the other colors) that are nearly accurate in the mean, though not precise. The equations of \citet{1994ApJ...422..158O} predict that $R_{VI,BV\rm{sfd}} = 0.249R_{V} + 0.562$ , where we have made sure to use the filter definitions of \citet{1998ApJ...500..525S} for calculating $E(B-V)$, while maintaining our definitions for $E(V-I)$. The mean value of $R_{VI,BV\rm{sfd}} = 1.179$ implies $R_{V}=2.47$, which is consistent with the value of $R_{V} \approx 2.5$ inferred later in this work. 

However, the values of $E(B-V)_{\rm{SFD}}$ toward the bulge are not precise. The 15\% scatter in $R_{VI,BV\rm{sfd}}$ does not seem to correspond to any known reddening index. Its correlation with $E(J-K_{s})/E(V-I)$ is $\rho=-0.04$, and its correlation with $E(V-I)$ is $\rho=+0.02$, both negligible values which undermine the prospect for an extinction-based explanation to the dispersion in $R_{VI,BV\rm{sfd}}$.


\section{The Reddening Law Toward the Bulge is Non-Standard and Non-Uniform}
\label{sec:ReddeningLawEstimates}
There are two broadly used methods to convert a measurement of reddening into a derived extinction. The first is to assume a universal total-to-selective extinction ratio, i.e.:
\begin{equation}
A_{I}   =  \frac{A_{I}}{E(V-I)}{\times}E(V-I) = R_{I}{\times}E(V-I),
\end{equation}
where we showed in Section \ref{Sec:ReddeningTheory} that $R_{I} \approx 1.45$ if one assumes standard extinction.

\begin{figure}[H]
\begin{center}
\includegraphics[totalheight=0.37\textheight]{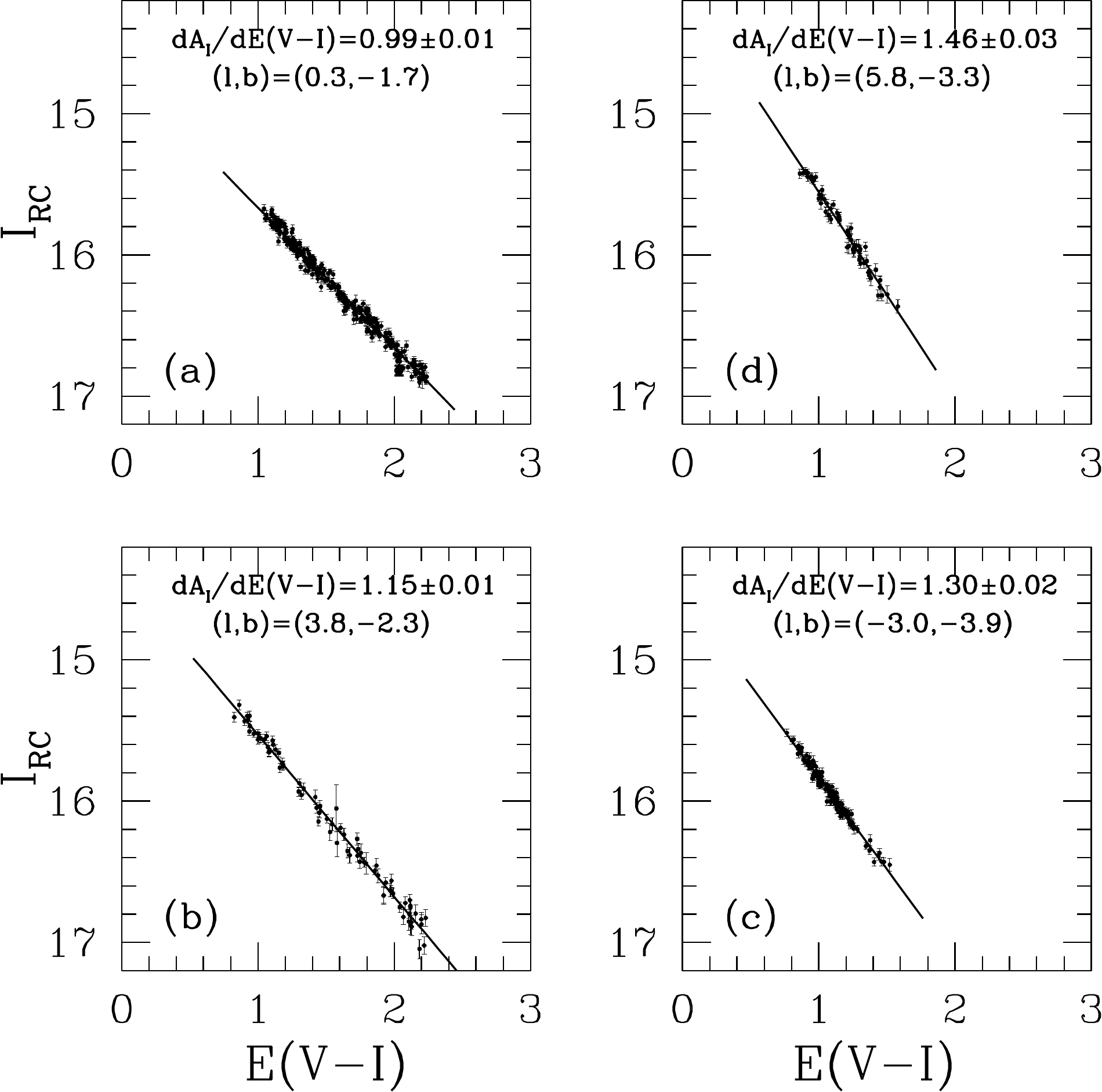}
\end{center}
\caption{We show the scatter plots of $I_{RC}$ vs $E(V-I)$ for four of the directions measured in this section, as well as the derived value of $dA_{I}/dE(V-I)$ with error. Plots are organized such that $dA_{I}/dE(V-I)$ appears in ascending order, counter-clockwise from TOP-LEFT.} 
\label{Fig:LocalReddening3Mosaic2}
\end{figure}

The second method is to compute the linear regression of magnitude as a function of color, and to infer the total-to-selective extinction ratio in that manner:
\begin{equation}
R_{I} = \frac{A_{I}}{E(V-I)} = \frac{dA_{I}}{dE(V-I)}.
\label{EQ:RIassumption}
\end{equation}
This method has the desirable characteristic that it is independent of any assumption of the intrinsic luminosity of whichever standard candle is being used: it depends only on the differential colors and magnitudes between sightlines, a robust quantity. This method has been used to infer the optical reddening toward the bulge \citep{2003ApJ...590..284U,2004MNRAS.349..193S,2010A&A...515A..49R,2012ApJ...750..169P}, the near-IR reddening toward the bulge \citep{2006ApJ...638..839N,2008ApJ...680.1174N,2009ApJ...696.1407N}, and to calibrate the (near+mid)-IR reddening in the disk \citep{2005ApJ...619..931I,2009ApJ...707..510Z,2009ApJ...707...89G}, where the latter used the analogous relation for color-color regressions rather than magnitude-color regressions. The assumption of Equation (\ref{EQ:RIassumption}) is that the functional dependence of the variation in extinction on the variation in reddening is equal to the total-to-selective extinction ratio. The $\sim$100 deg$^{2}$ Galactic bulge component of the OGLE-III survey allows us to ascertain whether or not this assumption is valid. Four of the fits to $dA_{I}/dE(V-I)$ are shown in Figure \ref{Fig:LocalReddening3Mosaic2}, visually demonstrating that the variability of the reddening law is a robust result.

We measure $dA_{I}/dE(V-I)$ as a function of position as follows. We keep the 9,014 RC measurements qualified as reliable in Section \ref{sec:RCmeasurements}, but remove those toward double-RC sightlines: $( |l + 0.5| \leq 4.5^{\circ})$ and $(|b| \geq 4.75^{\circ})$. We then measure, for the 1,690 OGLE-III subfields used in this work that are not toward a double-RC sightline, the linear regression of $(I_{RC} + 0.03(l-l_{\rm{central}}))$ vs $(V-I)_{RC}$ of every RC centroid measurement within 30$\arcmin$ of the subfield central coordinate. The factor of $0.03(l-l_{\rm{central}})$ is a first-order correction for the Galactic bulge's orientation which negligibly impacts the final values. The distribution of $dA_{I}/dE(V-I)$ as a function of direction is plotted in the bottom panel of Figure \ref{Fig:Mapper17D4Paper4Maps2}. Several results are immediately evident. We confirm the finding that $dA_{I}/dE(V-I)$ is smaller toward the inner Galaxy, as we obtain a mean value of $dA_{I}/dE(V-I)=1.215$. Also evident, however, is that for sightlines further separated from the minor axis we see higher, more standard values of $dA_{I}/dE(V-I)$, a significant number of sightlines with $l \geq 5^{\circ}$ are color-coded black, meaning  $dA_{I}/dE(V-I) \geq 1.34$. No Galactic bulge fields with values of  $dA_{I}/dE(V-I)$ so close to standard were found in the OGLE-II investigations of \citet{2003ApJ...590..284U} and \citet{2004MNRAS.349..193S}, as these were all closer to the Galactic plane and/or to the Galactic minor axis, and thus more heavily probed the extinction properties of the inner Galaxy. These values were also not found by the investigation of \citet{2012ApJ...750..169P}, as OGLE-III had lower cadence toward sightlines further from the Galactic minor axis.  The trend toward more ``standard'' reddening properties for sightlines further from the Galactic center may be linked to the finding of \citet{2009ApJ...707..510Z}, who found that the functional dependence of Galactic extinction in the mid-IR depends on Galactocentric radius, with shallower (greyer) values measured toward the inner Galaxy.

\begin{figure}[H]
\begin{center}
\includegraphics[totalheight=0.3\textheight]{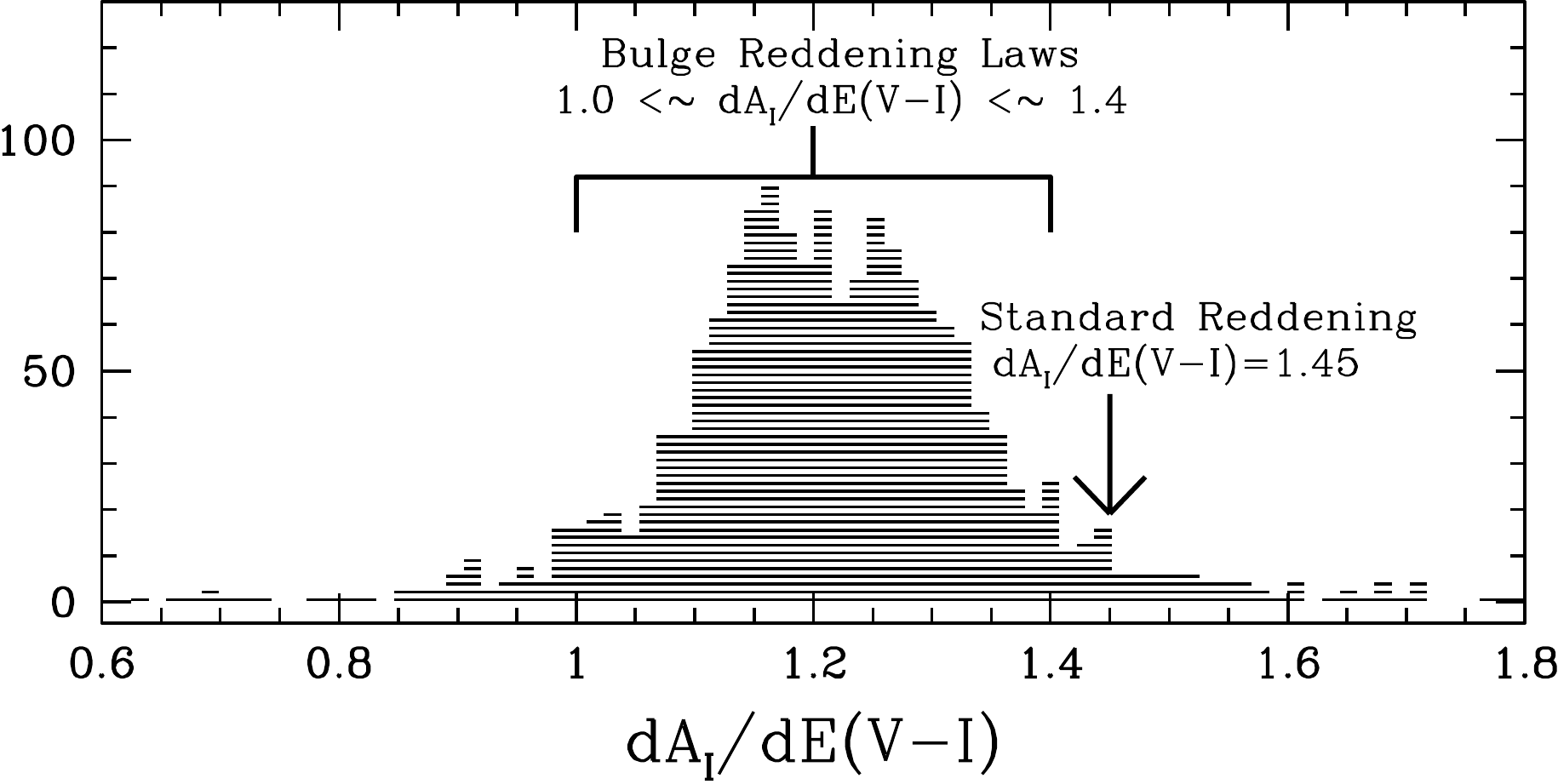}
\end{center}
\caption{Distribution of measured $dA_{I}/dE(V-I)$ values. The mean is 1.215 and the standard deviation is 0.18, with a mean error of 0.09. As such, nearly all sight lines investigated in this work have a smaller total-to-selective extinction ratios than the standard reddening law of $dA_{I}/dE(V-I)=1.45$. } 
\label{Fig:Mapper17RI}
\end{figure}

\begin{figure}[H]
\begin{center}
\includegraphics[totalheight=0.37\textheight]{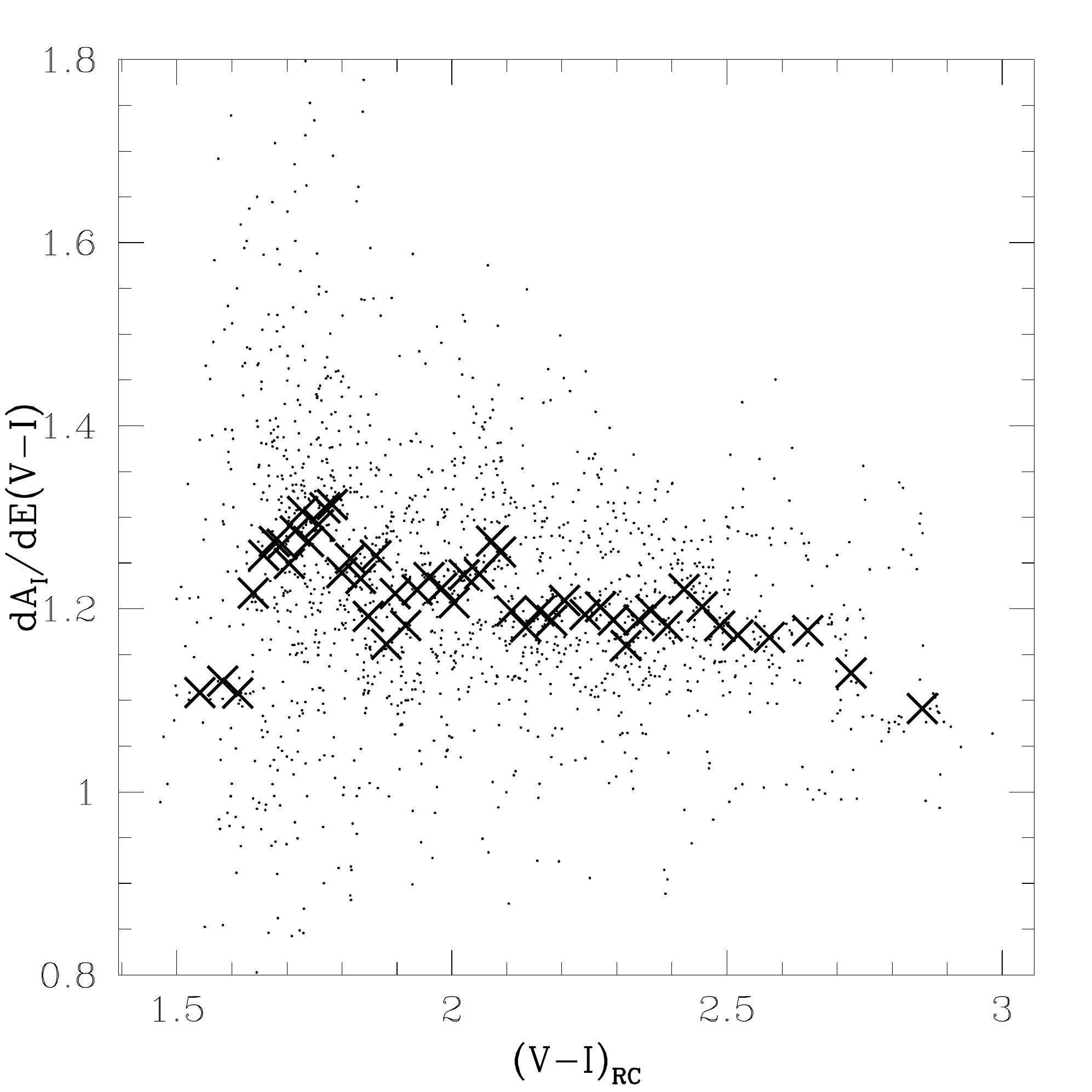}
\end{center}
\caption{Scatter plot of $dA_{I}/dE(V-I)$ versus the mean $(V-I)_{RC}$ of the RC centroids from which $dA_{I}/dE(V-I)$ is measured. Thick black X's denote the medians to 50 bins. } 
\label{Fig:RIanalysis2}
\end{figure}

The histogram of $dA_{I}/dE(V-I)$ is shown in Figure \ref{Fig:Mapper17RI}. We plot the distribution of $dA_{I}/dE(V-I)$ as a function of mean $(V-I)_{RC}$ in Figure \ref{Fig:RIanalysis2}. The fact that the reddening law is smaller than the local value of $R_{I} \approx 1.45$ is itself independent of the reddening. However, there is a small trend between the two variables: we also measure a small correlation between $dA_{I}/dE(V-I)$ and $(V-I)_{RC}$ of $\rho=-0.11$. This correlation may be due to the fact that reddening towards the plane is on average both higher and more heavily-dependent on the properties of dust further towards the inner Galaxy.

\subsection{Attempt at Deriving the Extinction From a Single Reddening Value}
\label{subsec:SingleReddening}
\begin{figure}
\begin{center}
\includegraphics[totalheight=0.75\textheight]{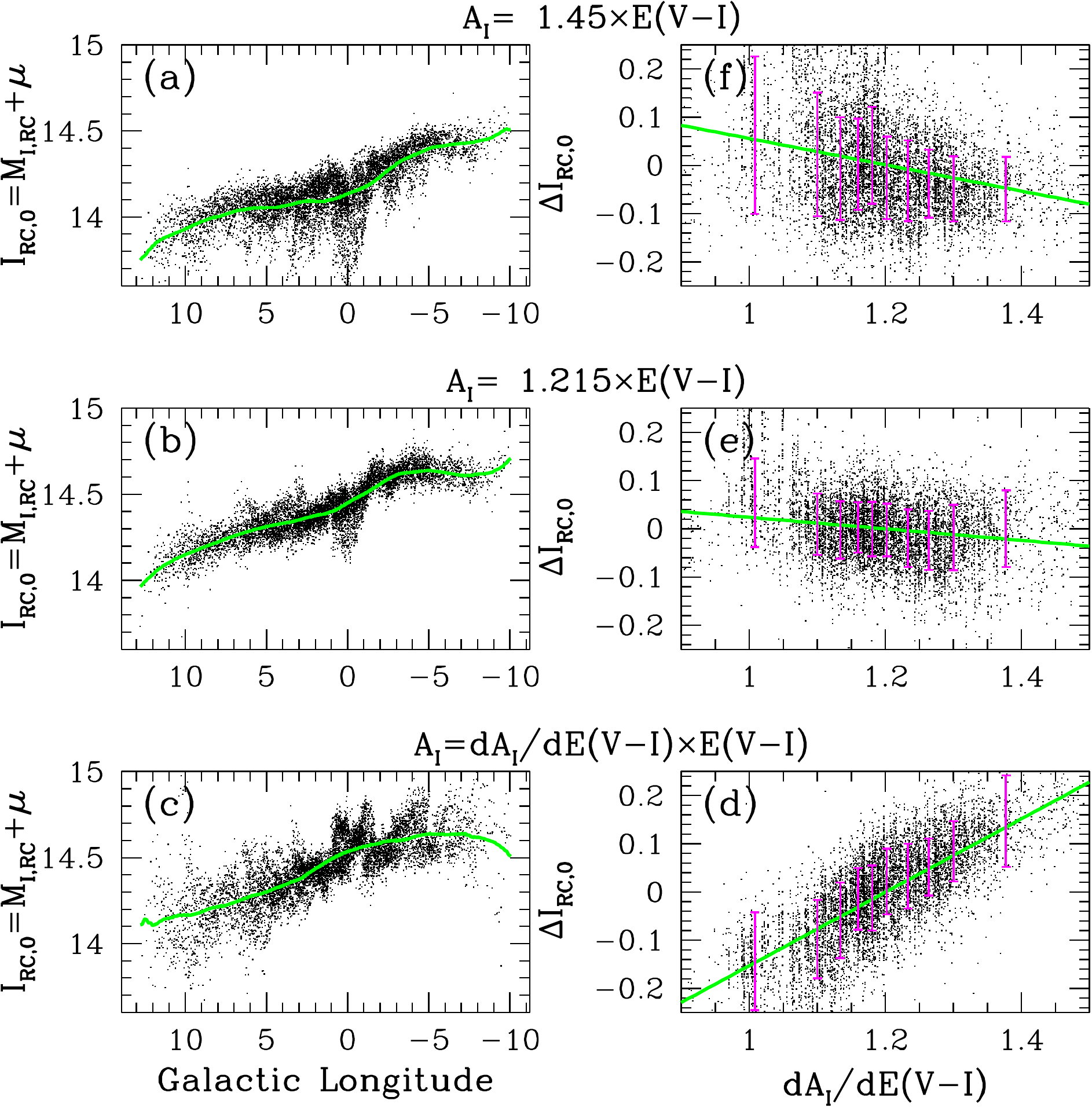}
\end{center}
\caption{\textbf{Panels (a,b,c)} The scatter of $\sim$9,000 dereddened RC magnitude, $I_{RC,0} = M_{I,RC}+\mu$ as a function of longitude for three different assumptions of the reddening law. Green line shows a moving linear fit to the points.  \textbf{Panels (d,e,f)} The residuals of $I_{RC,0}$, ${\Delta}I_{RC,0} = I_{RC,0,\rm{Fit}} - I_{RC,0}$ as a function of $dA_{I}/dE(V-I)$, where the latter is computed using 30$\arcmin$ circles. Error bars denote the dispersion of ${\Delta}I_{RC,0}$ in each bin. } 
\label{Fig:LongMagMosaic3}
\end{figure}

We attempt to derive the extinction to the Galactic bulge using three different methods that might appear valid.  The three methods are to assume that:
\begin{enumerate}
 \item $R_{I}=1.45$ everywhere, the ``standard'' value of $R_{I}$ when evaluating the extinction fits of  \citet{1989ApJ...345..245C} and \citet{1994ApJ...422..158O} at the value $R_{V}=3.1$;
 \item $R_{I}=1.215$ everywhere (the mean value found in the previous section);
 \item $A_{I}/E(V-I) = dA_{I}/dE(V-I)$, as per Equation (\ref{EQ:RIassumption}).
\end{enumerate}
We plot the resultant values of $I_{RC,0}=M_{I,RC}+\mu$ as a function of longitude on the left panels of Figure \ref{Fig:LongMagMosaic3}. On the right panels we plot the residuals of $I_{RC,0}$ relative to a moving fit as function of $dA_{I}/dE(V-I)$. It is clear from Figure \ref{Fig:LongMagMosaic3} that each of the three methods predicts large variations in the dereddened magnitude of the RC toward the same longitude in the Galaxy, which we will subsequently demonstrate to be a sign of failure.

The failure of the first method, evident in the top panels of Figure \ref{Fig:LongMagMosaic3}, is the least surprising. It not only fails, it fails spectacularly. Huge structures are present, stretching downwards from the bulge at every longitude, contributing to a scatter of 0.110 mag. At $l=0^{\circ}$, the best-fit value of $I_{RC,0}$ is 14.131 mag. If one assumes $M_{I,RC}=-0.12$ (Section \ref{sec:Tuc6791}), the implied distance to the Galactic center is $\sim$7.0 kpc, very much on the low side. The correlation coefficient between $dA_{I}/dE(V-I)$ and $I_{RC,0}$ is $\rho=-0.32$, demonstrating that there is information in $dA_{I}/dE(V-I)$ not used by this method that could improve the fit, as the dereddened distance modulus should not be sharply sensitive to the reddening law. Though the non-standard $VI$ extinction toward the bulge is already firmly established in the literature \citep{2001ApJ...547..590G,2003ApJ...590..284U,2004MNRAS.349..193S,2012ApJ...750..169P}, we further demonstrate it here as it is a critical point worthy of reiteration. 

The second method does much better. The scatter is the lowest of the three methods, at 0.070 mag per point.  The $\sim$36\% reduction in scatter means that the higher value of $R_{I}$ demanded by a standard reddening law was itself contributing 0.085 mag to the scatter. The implied distance to the Galactic center is a reasonable 8.19 kpc. However, the method still has some problems. The correlation coefficient between $dA_{I}/dE(V-I)$ and $I_{RC,0}$ is $\rho=-0.23$, demonstrating that there remains pertinent information in $dA_{I}/dE(V-I)$ that could be used to improve the fit.  There are various unphysical features in the plot, including a streak of points extending downwards near $l=0^{\circ}$, and two streaks extending upwards at $l=+3^{\circ},+6^{\circ}$.

Perhaps the most surprising result of this section is that the assumption that $A_{I}/E(V-I) = dA_{I}/dE(V-I)$, shown in the bottom two panels, does in fact not work. There are huge features that appear as an unphysical zigzag pattern superimposed on the Galactic bulge. The correlation coefficient between $dA_{I}/dE(V-I)$ and $I_{RC,0}$ is $\rho=+0.78$. Its absolute value is more than twice as large as that from the other two methods. However, the correlation coefficient has the opposite sign, further confirming that there is information in  $dA_{I}/dE(V-I)$ even if it is not yet clear what that information is.

That $A_{I}/E(V-I) \neq dA_{I}/dE(V-I)$ is a shocking result. These systematics prove that the functional dependence of the variation in extinction on the variation in reddening \textit{is not equal} to the total-to-selective extinction ratio. This assumption is often used in the literature without even stating that it is an assumption, demonstrating the extent to which this assumption appears natural. We understand that many readers will be skeptical about this point -- it is not unreasonable to look at the line fits in Figure \ref{Fig:LocalReddening3Mosaic2}, to see how well they go through the points, and to conclude ``that has to be extinction law''. We dedicate the following subsection to the requirement of providing a heuristic explanation to this phenomenon. 

\subsection{Variations in the Reddening Law Along the Line of Sight and Composite Extinction Bias}
\label{subset:composite}
The extinction law is a variable function of direction. Given that, we expect that it should vary within sightlines, and that the total extinction along a sightline to a distance $R$ will therefore be an integral:
\begin{equation}
A_{I} =  \int_0^R \frac{dA_{I}}{dE(V-I)}(r)   \frac{dE(V-I)}{dr}(r)  \,dr,
\label{EQ:TrueExtinction}
\end{equation}
where $({dA_{I}}/{dE(V-I)})(r)$ and $({dE(V-I)}/{dr})(r)$ are the reddening law and reddening density as a function of radius. It logically follows from Equation \ref{EQ:TrueExtinction} that the true extinction law for a given sightline will be an average weighted by the total reddening contributed from each type of interstellar medium and corresponding dust properties intersecting the line of sight. 

The slope $dA_{I}/dE(V-I)$ has historically been estimated by measuring the functional dependence of the variation in extinction on the variation in reddening. However, given that there will be many different extinction laws along a line of sight, that need not be the case. Consider a two-point regression, measured from RC centroids that are 30$\arcmin$ apart on the sky. At a distance of 1 kpc, the two sightlines will be $\sim$8 pc apart. At a distance of 4 kpc, the two sightlines will be $\sim$30 pc apart. The difference in extinction between sightlines towards the bulge will therefore be statistically biased toward whichever kind of dust is to be found deeper into the inner Galaxy, as distinct sightlines diverge linearly with distance. $dA_{I}/dE(V-I)$ is a weighted average of different extinction laws, but the weights are not the same weights as those which go into Equation \ref{EQ:TrueExtinction}. The name we assign to this phenomenon is ``composite extinction bias'': The functional dependence of the variation in extinction on the variation in reddening is biased toward the kind of extinction which contributes differentially to different sightlines. A class of extinction that contributes equally to both sightlines will obviously contribute to the true value of $A_{I}/E(V-I)$ of both sightlines, but it will not contribute to the scatter in $A_{I}$ and $E(V-I)$, and thus not to $dA_{I}/dE(V-I)$. Though we discuss the bulge here, we expect that the same phenomenon may apply in any direction where the extinction law varies along the line of sight. 

\begin{figure}[H]
\begin{center}
\includegraphics[totalheight=0.4\textheight]{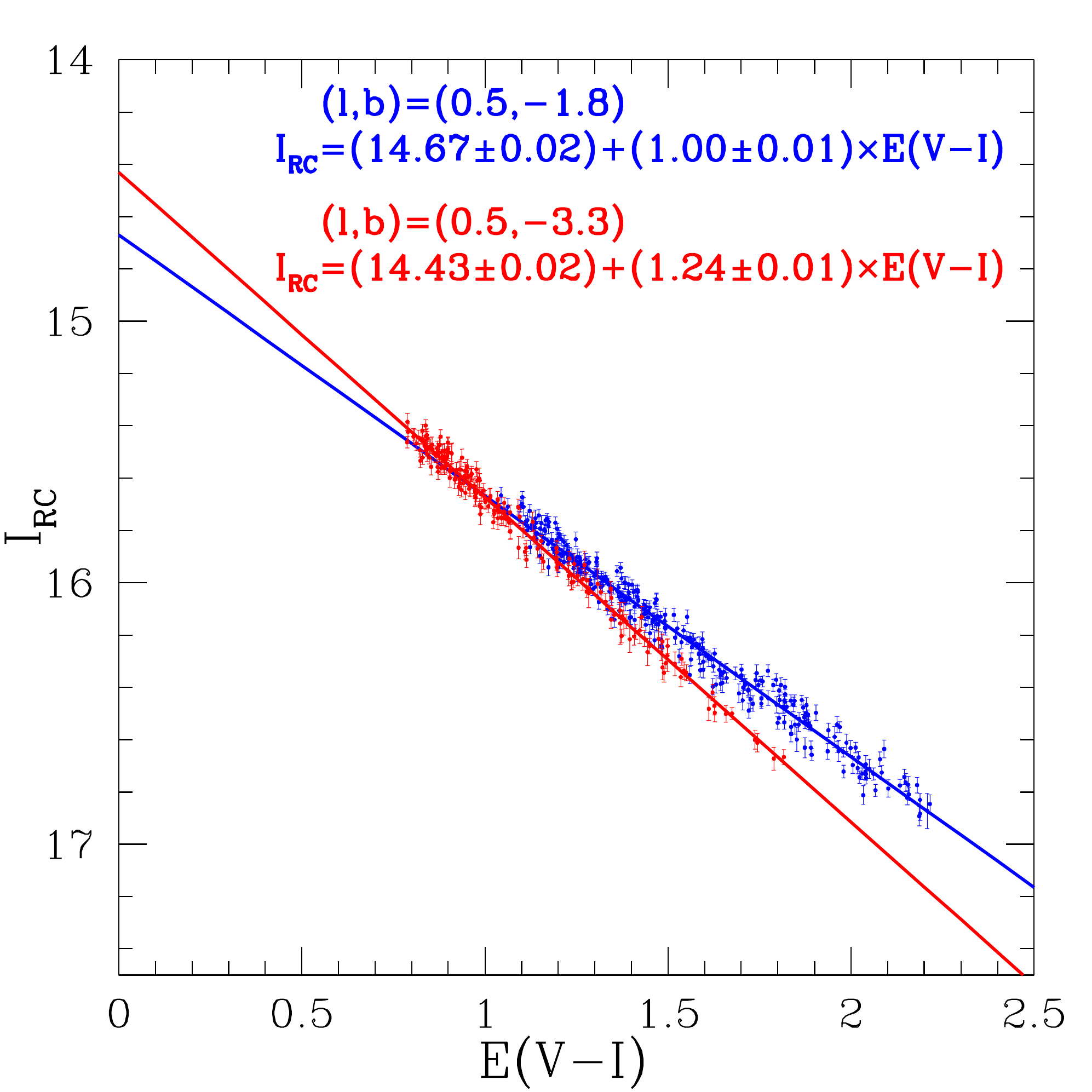}
\end{center}
\caption{Scatter of $I_{RC}$ vs $(V-I)_{RC}$ near two sightlines. The measured reddening laws $dA_{I}/dE(V-I)$ are clearly different, but extrapolating the reddening laws back to $E(V-I)=0$ leads to an unphysical difference in the value of $I_{RC,0}+\mu = 0.24$ mag. We thus conclude that the slope  $dA_{I}/dE(V-I)$  will be different for lower values of reddening. } 
\label{Fig:LocalReddening3MosaicB}
\end{figure}

We show the derivations of $dA_{I}/dE(V-I)$ for two different directions in Figure \ref{Fig:LocalReddening3MosaicB}. These two fits yield different values of  $dA_{I}/dE(V-I)$, and values of $I_{RC,0}$ that differ by 0.24$\pm$0.03 mag. These two directions are at the same longitude, so the Galactic bulge's orientation should affect both directions equally, and thus the difference in $I_{RC,0}=\mu+M_{I,RC}$ cannot be due to a difference in distance modulus, as for a mean distance to the Bulge of 8.2 kpc (estimated later in this work), this offset implies a distance difference of $\sim$900 pc, some $\sim 4{\times}$ larger than  would be expected from latitude-dependent projection effects \citep{2007A&A...465..825C}. It also cannot plausibly be due to a difference in absolute magnitude, as RC stars have a metallicity-dependence of $\sim$0.2 mag dex$^{-1}$ in $I$ \citep{2001MNRAS.323..109G,2010AJ....140.1038P}, whereas from Equation \ref{EQ:colorterms3} and as such an unphysical mean metallicity of [Fe/H]$\approx$+1.20 would be required to understand the different intercept. The difference of 0.24$\pm$0.03 mag is thus seemingly unphysical, but it directly follows from extrapolating the fits in Figure \ref{Fig:LocalReddening3MosaicB} to $E(V-I)=0$. We thus conclude that the fits should not be extrapolated back to $E(V-I)=0$. There is a component of the extinction that contributes to the extinction of these sightlines, but much less strongly to the scatter of their extinctions. 

We are thus left with a dilemma. We have demonstrated that $dA_{I}/dE(V-I) \ne A_{I}/E(V-I)$, yet the correlations found in Section \ref{subsec:SingleReddening} demonstrate that it has some physical meaning. There is no obvious way to probe any screen of extinction that would contribute equally to neighboring sightlines without making use of problematic assumptions as to the intrinsic values of distance modulus. We conclude that the conversion from reddening to extinction cannot be done accurately with a single reddening value. The solution we propose is to utilize multiple reddening values, in different bandpasses. 

\section{Combining $E(V-I)$ and $E(J-K_{S})$ to Yield a Robust Extinction Estimate}
\label{sec:RJKVI}

\subsection{The Distribution of $E(J-K_{s})/E(V-I)$ }
\begin{figure}[H]
\begin{center}
\includegraphics[totalheight=0.4\textheight]{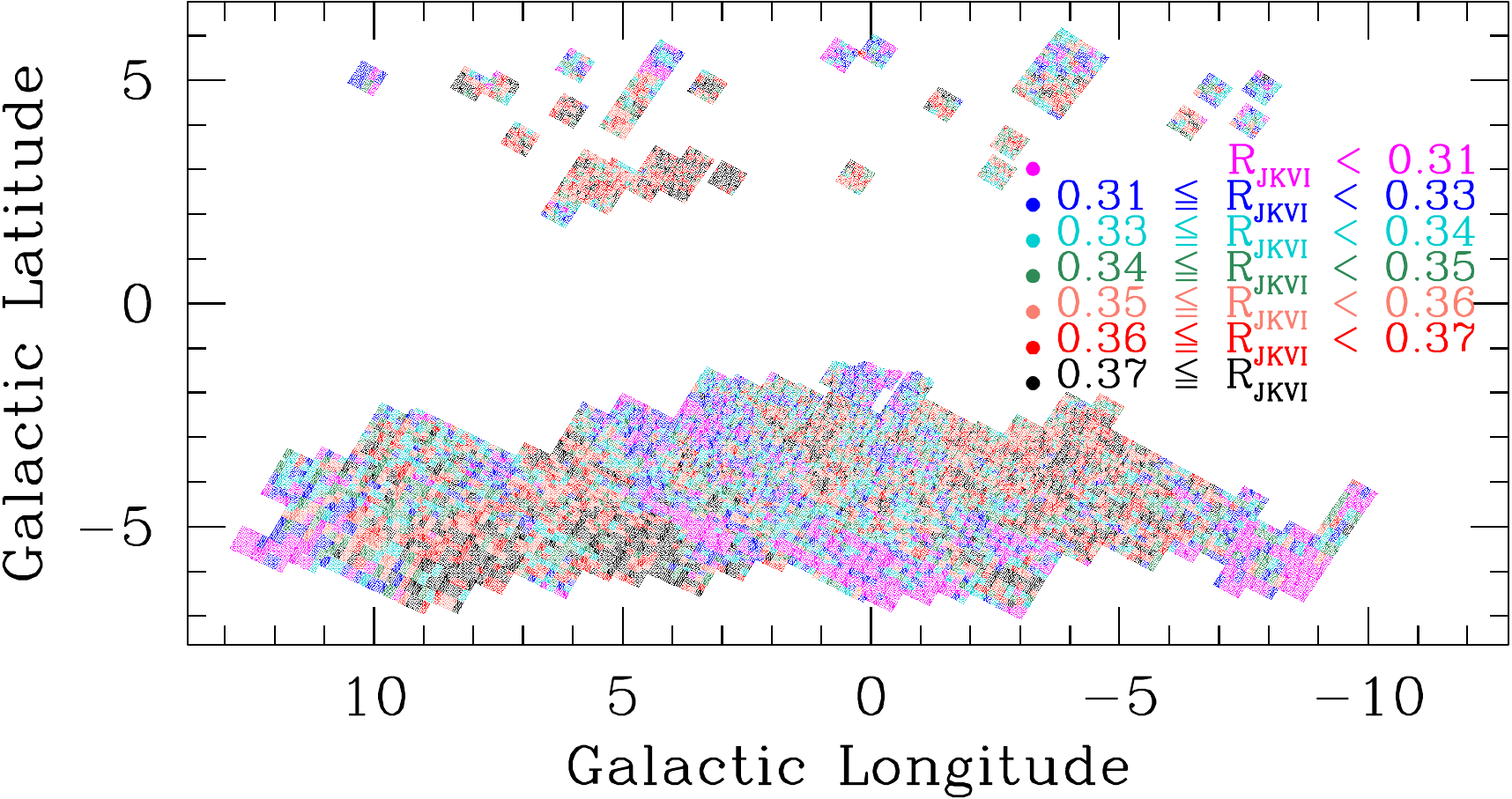}
\end{center}
\caption{Color map of the observed distribution of $R_{JKVI} = E(J-K_{s})/E(V-I)$. If the extinction curve toward the bulge were standard we would measure $R_{JKVI} \approx 0.41$, in contrast, even the demarcation between the 6th and 7th septiles, $R_{JKVI}=0.37$, is of a steeper extinction curve. The mean value, $R_{JKVI}=0.3433$, implies $R_{V} \approx 2.5$ \citep{1989ApJ...345..245C,1994ApJ...422..158O}. } 
\label{Fig:RIJKMapper}
\end{figure}

For each of our sightlines with a measured RC centroid we add a measurement of $E(J-K_{s})$, in the manner described in Section \ref{subsec:JMK2MASS}. We thus have two completely independent reddening measurements for every sightline: $E(V-I)$ and $E(J-K_{s})$. We plot the distribution of $R_{JKVI} = E(J-K_{s})/E(V-I)$ as a function of direction in Figure \ref{Fig:RIJKMapper}. The mean value of $R_{JKVI}$ is 0.3433, compared to $R_{JKVI} \approx 0.41$ if one assumes standard dust properties, as shown in Section \ref{Sec:ReddeningTheory}. Once again, the dust toward the inner Galaxy is characterized by a steeper extinction curve than standard.  Further from the minor axis, a reassuring similarity is observed between Figure \ref{Fig:RIJKMapper} and the bottom panel of Figure \ref{Fig:Mapper17D4Paper4Maps2}. Both have (relatively) more standard extinction further from the Galactic minor axis, toward approximately the same directions of $l=-4^{\circ},+5^{\circ}, b \approx -5^{\circ}$. 

We note two advantages of using $R_{JKVI}$ rather than $dA_{I}/dE(V-I)$. The first is that the ratio of two reddenings can be taken toward both single-RC and double-RC sightlines, as double-RCs do not differ in color \citep{2010ApJ...721L..28N,2010ApJ...724.1491M}. The second advantage is more fundamental: whereas the measurement of slopes $dA_{I}/dE(V-I)$ must be done over $\sim$30$\arcmin$ scales to be precise, independent values of $R_{JKVI}$ can be measured for each sightline, giving us very sharp resolution on the extinction law. The fact each of 9,014 measurements shown in Figure \ref{Fig:RIJKMapper} is independent of the other measurements confirms that the variations in the extinction law are real: if the variations were due to noise, they would not cluster together.

\subsection{$A_{I}$ As a Function of $E(V-I)$ and $E(J-K_{S})$}
\label{sec:AIdetermination}
The parameterizations of \citet{1989ApJ...345..245C} and \citet{1994ApJ...422..158O} assume that if one knows $R_{V}$, then one knows all the total-to-selective extinction ratios in the optical and near-IR. We do not know $R_{V}$ as we do not have measurements in $B$-band, however, knowing $R_{JKVI}$ is equivalent to knowing $R_{V}$ if one assumes the extinction law is a single-parameter function. Both \citet{1989ApJ...345..245C} and \citet{1994ApJ...422..158O} parameterize the extinction curve as $A_{\lambda} = A_{V}(a(\lambda) + b(\lambda)/R_{V})$, allowing us to obtain linear relations linking $A_{I}$, $E(V-I)$ and $E(J-K_{s})$. We invert the  equations of \citet{1989ApJ...345..245C} and find:
\begin{equation}
\frac{A_{I}}{E(V-I)} =  0.1713 + 3.078 {\times} \frac{E(J-K_{s})}{E(V-I)},
\end{equation}
which is equivalent to:
\begin{equation}
A_{I} = 1.228{\times}E(V-I)\biggl[1 + 2.507{\times}(R_{JKVI}-0.3433)   \biggl], 
\label{EQ:CardelliPrediction}
\end{equation}
whereas if we use the values of \citet{1994ApJ...422..158O}:
\begin{equation}
A_{I} = 1.266{\times}E(V-I)\biggl[1 + 2.323{\times}(R_{JKVI}-0.3433)   \biggl].
\label{EQ:ODonnellPrediction}
\end{equation}
There is a simple interpretation to
Equations (\ref{EQ:CardelliPrediction}) and (\ref{EQ:ODonnellPrediction}). They are first-order expansions to the reddening law in the expansion variable $R_{JKVI}$, with smaller values of $R_{JKVI}$ implying a steeper extinction curve. A steeper extinction curve is conventionally interpreted as being due to smaller dust grains \citep{2003ARA&A..41..241D}. As an example, for the sightline near Baade's window observed by \citet{2010ApJ...725L..19B}, toward $(l,b)=(+1.06^{\circ},-3.81^{\circ})$, $E(V-I)=0.67$ and $R_{JKVI}=0.351$, yielding $A_{I,\rm{Cardelli}}=0.84$ and $A_{I,\rm{O'Donnell}}=0.86$. 

\begin{figure}[H]
\begin{center}
\includegraphics[totalheight=0.7\textheight]{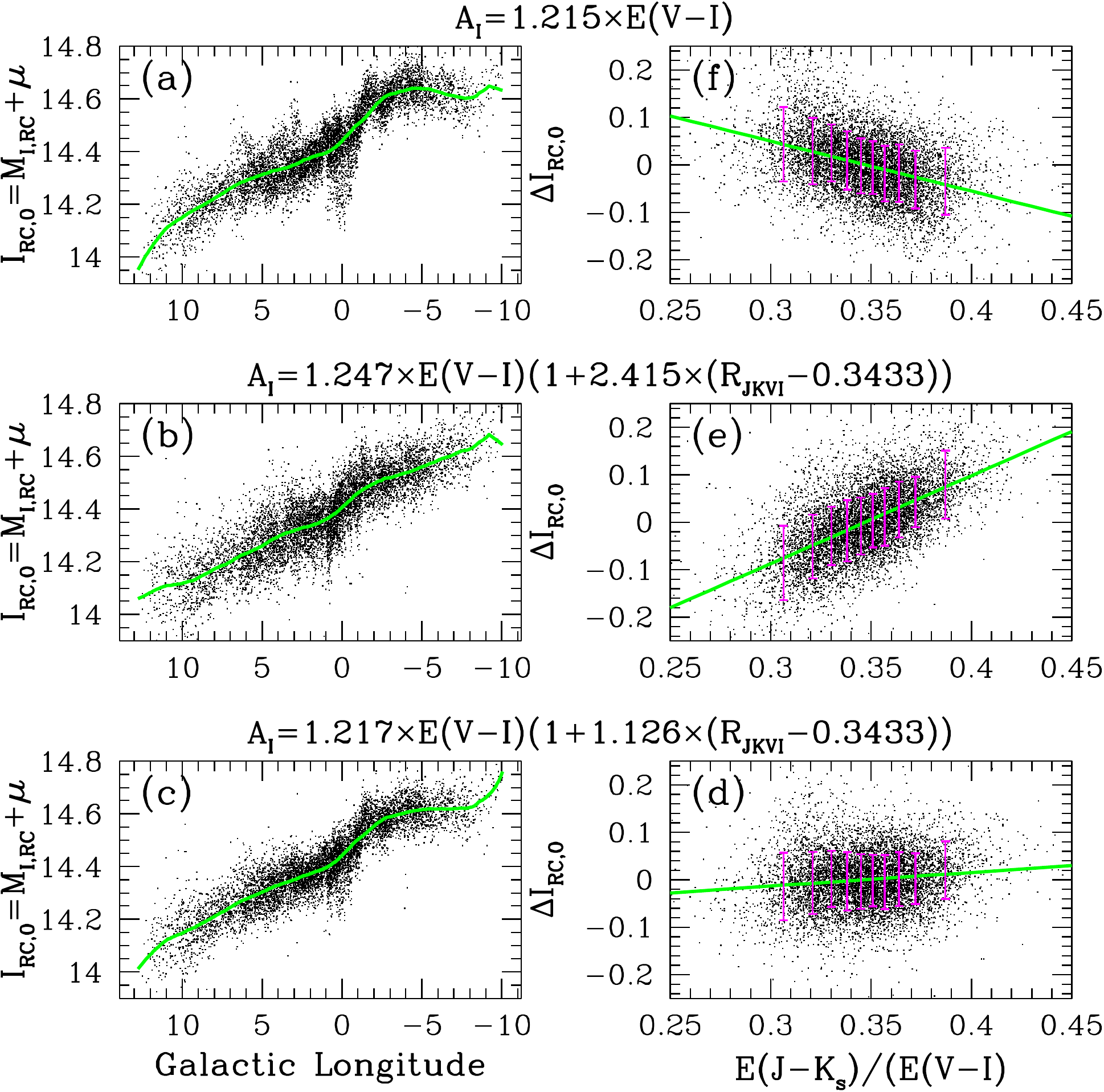}
\end{center}
\caption{\textbf{Panels (a,b,c)} The scatter of $\sim$9,000 dereddened RC magnitude, $I_{RC,0} = M_{I,RC}+\mu$ as a function of longitude for three different assumptions of the reddening law. Green line shows a moving linear fit to the points. \textbf{Panels (d,e,f)} The residuals of $I_{RC,0}$, ${\Delta}I_{RC,0} = I_{RC,0,\rm{Fit}} - I_{RC,0}$, as a function of $E(J-K_{s})/E(V-I)$. Error bars denote the dispersion of ${\Delta}I_{RC,0}$ in each bin. } 
\label{Fig:LongMagMosaic3B}
\end{figure}

Figure \ref{Fig:LongMagMosaic3B} shows the result of three different methods to fit for the extinction to the Galactic bulge. The top two panels show the result for $A_{I}=1.215{\times}E(V-I)$, with no information from $E(J-K_{s})$, as a comparison. The correlation between the residuals to the moving fit and the dereddened apparent magnitudes is $\rho=-0.39$, a large number demonstrating the failure of this method to use all the information available. The scatter is 0.070 mag.

In the middle panel, we plot the result given an extinction prescription that is the mean prediction of  \citet{1989ApJ...345..245C} and \citet{1994ApJ...422..158O}: $A_{I}=1.247{\times}E(V-I)(1+2.415{\times}(R_{JKVI}-0.3433))$. Surprisingly, this does less well than not using the information from $E(J-K_{s})$: the 1$\sigma$ scatter is increased to 0.078 mag. The correlation between the residuals to the moving fit and the dereddened apparent magnitudes is $\rho=+0.59$. That is nearly twice as large as that obtained when simply using $A_{I}=1.215E(V-I)$, though of the opposite sign. 

In the bottom two panels, we plot the results for the extinction prescription:
\begin{equation}
\begin{split}
 A_{I} &= 1.217{\times}E(V-I)\biggl[1+1.126{\times}(R_{JKVI}-0.3433)\biggl]  \\ 
       &= 0.7465{\times}E(V-I) + 1.3700{\times}E(J-K_{s}),  
\end{split}
\label{EQ:FinalReddeningLaw}
\end{equation}
values optimized by minimizing ${\chi}^2 = \Sigma[(I_{RC}-A_{I}-\overline{I_{RC}})/(\sigma_{I,RC}^2)]$, where $\overline{I,RC}$ is the mean dereddened RC magnitude within $1.5^{\circ}$ of longitude of that measurement, summed over all 9,014 reliable  measurements. We show the equation in two algebraic formats for clarity. This method produces the smallest scatter, at only 0.060 mag. The bottom-left panel shows the fewest structures. Finally, the correlation between the residuals to the moving fit and the dereddened apparent magnitudes is $\rho=+0.11$. It is the closest to zero of the three methods. That it is not exactly equal to zero may be due to the fact that the error on $I_{RC}$ is a function of location on the sky, and is therefore evidence that there are third parameters to the extinction law, the first two parameters being $E(V-I)$ and $R_{JKVI}$. These third parameters could be factors acting on only certain types of dust grains, or they could be phantom extinction parameters such as gradients in metallicity, age, or helium, which could easily masquerade as a variation in the reddening law.

Nevertheless, whatever these third parameters are, they are small. There are almost no features in the bottom-left panel of Figure \ref{Fig:LongMagMosaic3B}, unlike all other methods we previously have shown in this work. The scatter of 0.06 mag is very close to the noise floor: given our average errors of ${\sigma}_{I,RC}=0.034$ mag, ${\sigma}_{(V-I),RC}=0.01$ mag, and ${\sigma}_{(J-K_{s}),RC}=0.01$ mag, we could do no better than a dispersion of 0.039 mag. This noise floor does not include the impact of metallicity gradients and latitude-dependent projection effects. The method of Equation (\ref{EQ:FinalReddeningLaw}) yields a scatter of 0.06 mag, compared to 0.10 mag when assuming a standard reddening law of $A_{I}=1.45{\times}E(V-I)$. That is a factor 4 reduction in the variance once a conservative estimate of the noise floor is removed. Thus, to that high degree of accuracy, we have solved the longstanding observational challenge of the non-standard $VI$ extinction toward the Galactic bulge. 

\subsection{Caveats to the Reddening Law}
We discuss four caveats to the reddening law that can induce second-order effects on Galactic bulge studies at the $\sim$0.05 mag level. 

The first is that there is evidence for a dependence to the reddening law on the spectral energy distribution of stars being investigated. The reddening law found in this investigation is a bit shallower than that reported by \citet{2012ApJ...750..169P}, who used RR Lyrae stars as standard candles. This is partly because OGLE-III had lower cadence toward sightlines further from the minor axis, where the reddening law is relatively more standard. However, even once that is accounted for there is still a small effect. We compare RRab stars to the closest of the clean RC centroids described in Section \ref{sec:ReddeningLawEstimates} that are also not toward a double-RC, and that are within 5$\arcmin$ of an RRab star. We find:
\begin{equation}
(V-I)_{RC}-(V-I)_{RRab} = 0.552  -0.084{\times}((V-I)_{RC}-2.22) -0.83{\times}(|b|_{RC} - |b|_{RRab} - 0.021),
\label{EQ:RRRCcomparison}
\end{equation}
The latitude term captures the fact that an RRab star will be more reddened than the nearest RC centroid if it is closer to the plane. No significant trend is found for $I_{RC}-I_{RRab}$ regardless of whether or not we fit for a longitudinal term, so the effect is purely in $V$-band. This temperature-dependence is several times larger than that predicted from theory in Section \ref{Sec:ReddeningTheory}. The origin of the significance to the terms in Equation \ref{EQ:RRRCcomparison} is as such unresolved.

Metallicity gradients could have a small impact. From Section \ref{sec:Calibration}, we expect that for a 0.2 dex$^{-1}$ increase in metallicity at fixed age, from [M/H]$=0$ to [M/H]$=+0.20$, $M_{I,RC}$ should get fainter by 0.04 mag, and $(V-I)_{RC}$ redder by 0.06 mag, thus making $(J-K_{s})_{RC}$ redder by 0.04 mag. A 0.2 dex increase in the metallicity would therefore cause ``dereddened'' RC stars to appear \textit{brighter} by $\sim$0.03 mag, as the effect of overestimating the extinction (due to redder intrinsic colors) would be larger than that of the dimming of the RC in $I$. The actual amplitude will be modified depending on how [Fe/H] correlates to [$\alpha$/Fe], age, helium abundance, and binarity. The bulge vertical metallicity gradient measured at large separations from the plane is $\sim$0.6 dex kpc$^{-1}$  \citep{2008A&A...486..177Z,2011ApJ...732..108J}, though evidence  suggests this flattens out within 4$^{\circ}$ of the plane \citep{2000AJ....120..833R,2011A&A...534A...3G,2012ApJ...746...59R}. An additional source of uncertainty is that the mapping of other abundances onto [Fe/H] may depend on kinematics \citep{2012ApJ...749..175J}, and the Galactic bulge age-helium-metallicity relationship is only loosely constrained at high metallicities \citep{2012ApJ...751L..39N}. Once a clear picture of the bulge metallicity gradient emerges, in both latitude and longitude, there may be a need for an additional iteration to reddening and extinction maps.

\begin{figure}[H]
\begin{center}
\includegraphics[totalheight=0.40\textheight]{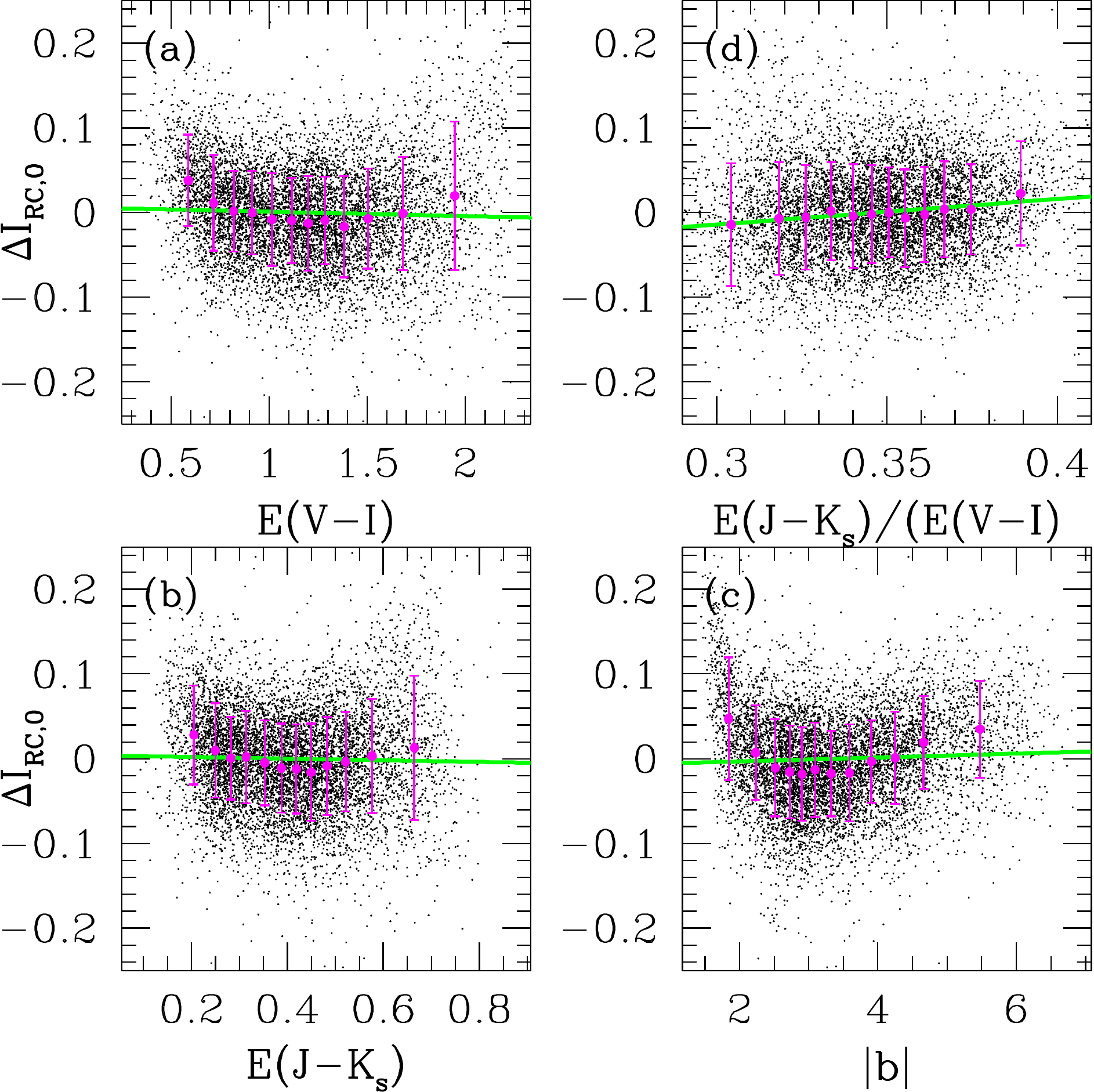}
\end{center}
\caption{The scatter in ${\Delta}I_{RC,0} = I_{RC,0,\rm{Fit}} - I_{RC,0}$, as a function of four parameters. There are no strong trends with $E(V-I)$, $E(J-K_{s})$ or $E(J-K_{s})/E(V-I)$. However, there are clear, non-linear trends with separation from the plane. For each panel, error bars denote the dispersion of ${\Delta}I_{RC,0}$ in each bin. } 
\label{Fig:LongMagMosaic2B}
\end{figure}

There appears to be a latitude term in the residuals to the extinction law when using Equation (\ref{EQ:FinalReddeningLaw}). We show the scatter in $I_{RC,0}$ as a function of four parameters in Figure \ref{Fig:LongMagMosaic2B}. Whatever trend there may be with $E(V-I)$, $E(J-K_{s})$ or $E(J-K_{s})/E(V-I)$ is no greater than $\sim$0.02 mag over the bulk of the data. However, there is a large, non-linear trend with separation from the plane, represented as the absolute value of the latitude. At large separations from the plane, $I_{RC,0}$ is $\sim$0.05 mag brighter than the faintest $I_{RC,0}$ values seen, at around $|b|=3^{\circ}$. This can plausibly be due to projection effects; at large separations from the plane, the distance to the maximum density of stars along a line of sight through a Galactic bulge becomes systematically smaller than distance to  the major-axis \citep{2007A&A...465..825C}. If that is the explanation, the same offset will show up in studies of the RC in $K_{s}$ and in mid-IR filters. However, for $|b| \lesssim 2^{\circ}$, we see another very large increase, where the RC becomes up to $\sim$0.1 mag brighter than its mean longitudinal value. Further investigation of this feature is warranted. An additional parameter to the extinction law may seem the most likely culprit, but variations in the metallicity or projection effects could also contribute. An extension of our methodology to incorporate additional reddenings, such as $E(B-V)$, $E(I-J)$, or mid-IR reddenings could help disentangle whether the remaining scatter is due to additional parameters of the properties of the interstellar medium, or due to the intrinsic properties of the bulge stellar population. 

\begin{figure}[H]
\begin{center}
\includegraphics[totalheight=0.46\textheight]{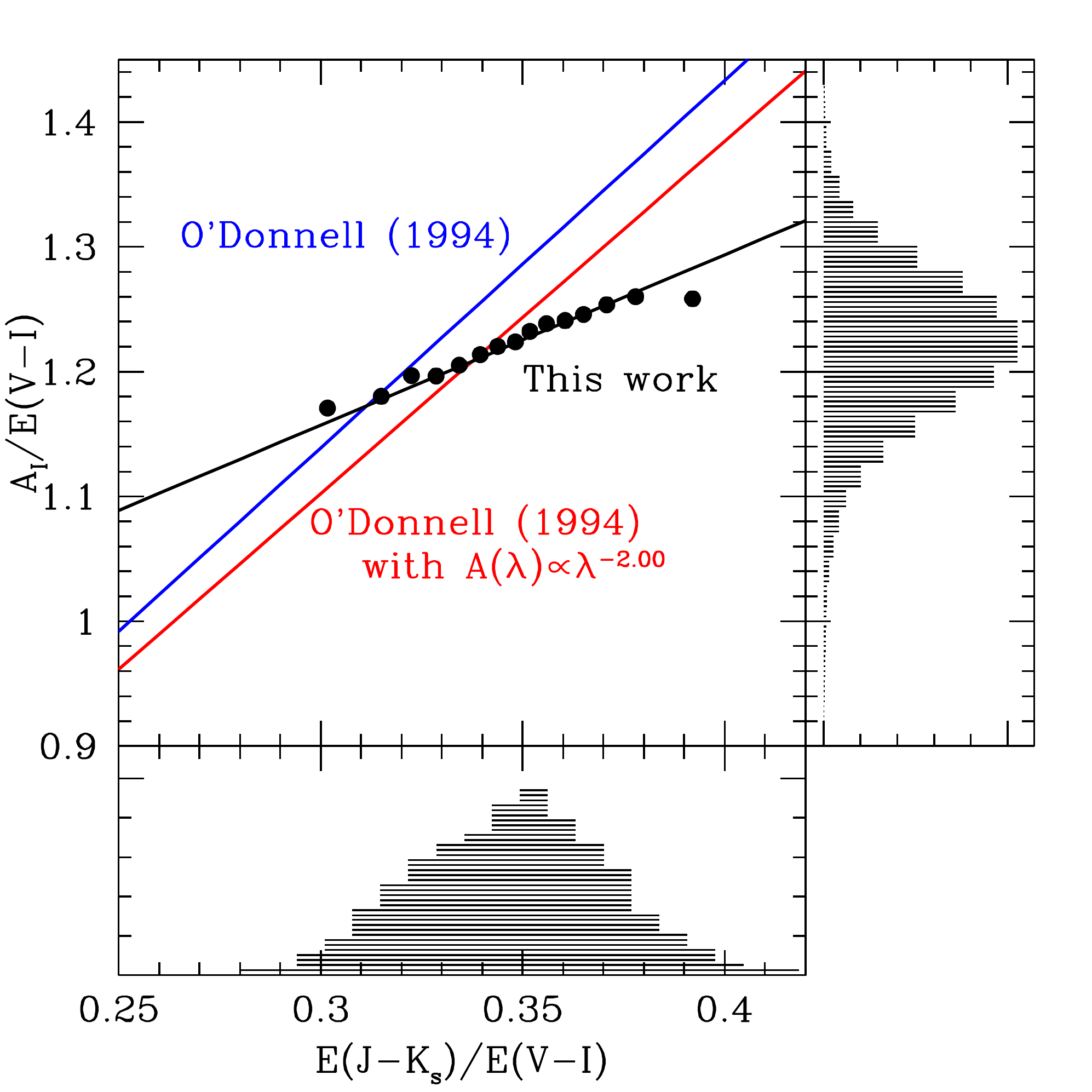}
\end{center}
\caption{The predicted relationship for $A_{I}/E(V-I)$ vs $E(J-K_{s})/E(V-I)$ using the equations of \citet{1994ApJ...422..158O} (blue) as written, using the equations of \citet{1994ApJ...422..158O} modified to assume $A(\lambda) \propto \lambda^{-2.00}$ to be consistent with the result of \citet{2009ApJ...696.1407N} (red), and the best-fit relationship found in this work (black), with binned medians shown as black circles. The red and black curves intersect at $(E(J-K_{s})/E(V-I), A_{I}/E(V-I)) = (0.337, 1.208)$, very close to the respective mean values of 0.3433 and 1.217 found in this work. Histograms of $E(J-K_{s})/E(V-I)$ and $A_{I}/E(V-I)$ values overplotted.} 
\label{Fig:ReddeningLaws}
\end{figure}

Finally, we note what may be the most important caveat to the reddening law. Our mean reddening values, $A_{I}/E(V-I)=1.22$ and $R_{JKVI} = 0.3433$, both suggest an $R_{V} \approx 2.5$ extinction curve. However, this is done without any actual $B$-band measurements, and thus the inference could be affected by systematics. The formalism of \citet{1989ApJ...345..245C} and \citet{1994ApJ...422..158O} both assume a universal extinction law in the near-IR, $A(\lambda) \propto \lambda^{-1.61}$, that is independent of $R_{V}$. However, it has been demonstrated that the near-IR reddening law toward the inner Galaxy is steeper. \citet{2009ApJ...696.1407N} used RC stars to measure $A(\lambda) \propto \lambda^{-2.00}$, a finding supported by \citet{2011ApJ...737...73F}, who used line-emission to study the extinction curve.

It may be that the exponential slope of the near-IR extinction law rises to higher values for lower values of $R_{V}$. \citet{1989ApJ...345..245C} and \citet{1994ApJ...422..158O} would not have been able to measure this, as none of their 51 sightlines probed values of $R_{V}$ as low as the mean value toward the bulge inferred in this work. This could explain why the coefficients to Equation (\ref{EQ:FinalReddeningLaw}) do not match the predicted values. In Figure \ref{Fig:ReddeningLaws}, we show that adjusting the equations of \citet{1994ApJ...422..158O} to assume $A(\lambda) \propto \lambda^{-2.00}$ yields predictions more consistent with our measurements for the typical values of $E(J-K_{s})/E(V-I)$ found in this work. This may be a coincidence, but it would be worth investigating, once data in more bandpasses become available, whether the exponential slope of the near-IR extinction curve rises as $R_{V}$ goes down. We note that studies of the extinction in the hosts of type Ia SNe routinely find an $R_{V} \approx 2.5$ extinction curve \citep{2010A&A...523A...7G,2011A&A...529L...4C,2011ApJ...731..120M}, and then subsequently assume the functional dependence on wavelength calibrated in investigations of Milky Way extinction. If that functional form were modified, it would have significant implications not just for Galactic studies, but also for cosmology.

\section{Have Studies of RR Lyrae in MACHO Photometry Constrained the Reddening Law Toward the Bulge?}
\label{sec:MACHO}
\citet{2008AJ....135..631K} investigated the reddening law toward Galactic bulge RR Lyrae stars observed by the MACHO survey. They inferred a reddening law of $R_{V,VR} = A_{V}/E(V-R)=4.3 \pm 0.2$, consistent with the standard extinction curve resulting from $R_{V}=3.1$. It is perhaps the most significant investigation in the literature arguing for a standard reddening law toward the Galactic bulge\footnote{An analysis of 16 planetary nebulae in the direction of the bulge by \citet{2013A&A...550A..35P} was posted to astro-ph on January 16th, 2013, after this investigation was submitted for publication. The authors argue for a standard  $R_{V}=3.1$ reddening law for the bulge.} and thus needs to be understood. We demonstrate here that this result is likely to be a consequence of an incorrect assumption: that the transformations of \citet{1999PASP..111.1539A} correctly calibrate the MACHO $V$ and $R$ filters into Johnson $V$ and Kron-Cousins $R$ filters. 

\begin{figure}[H]
\begin{center}
\includegraphics[totalheight=0.52\textheight]{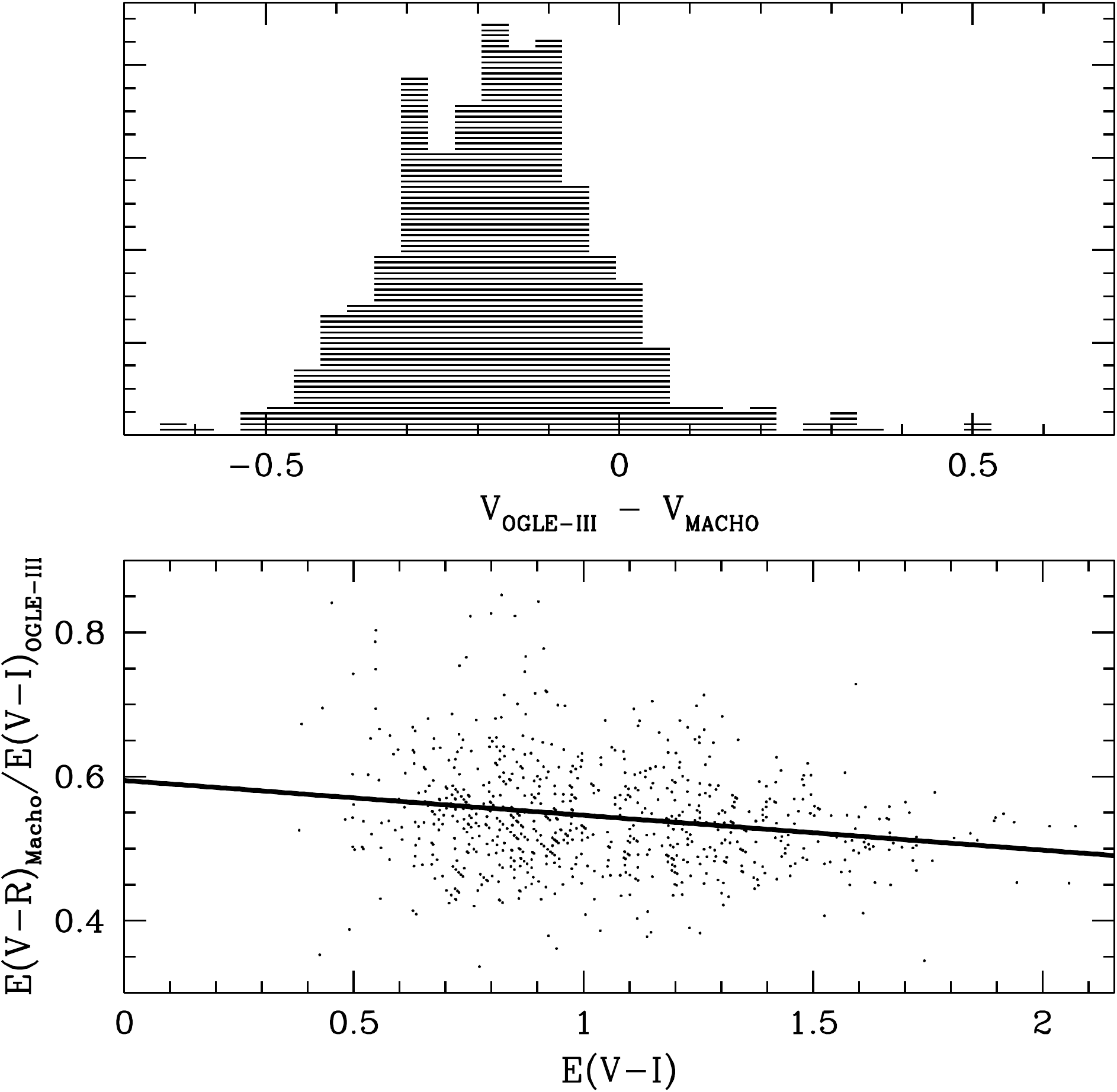}
\end{center}
\caption{TOP: Comparison of RR Lyrae stars from MACHO \citep{2008AJ....135..631K} with those from OGLE-III \citep{2011AcA....61....1S}. The mean magnitude of RR Lyrae stars in $V_{\rm{OGLE-III}}$ is 0.16 mag brighter than that of $V_{\rm{MACHO}}$. BOTTOM: $E(V-R)_{\rm{MACHO}} /E(V-I)_{\rm{OGLE-III}}$ as a function of E(V-I).  The mean value of $E(V-R)_{\rm{MACHO}} /E(V-I)_{\rm{OGLE-III}}=0.544$ is different from the value of   $E(V-R) /E(V-I)=0.42$ that would be expected if MACHO photometry was correctly calibrated on the Kron-Cousins system. } 
\label{Fig:RRmachoogle}
\end{figure}

We test this by first matching the MACHO RR Lyrae catalog of \citet{2008AJ....135..631K} and OGLE-III RR Lyrae catalog of \citet{2011AcA....61....1S}, keeping only the RRab stars, and keeping only those within 1$\arcmin$ of a reddening measurement (as measured in this work), for a cross-match of 788 sources. We plot the comparison in Figure \ref{Fig:RRmachoogle}. In the top-panel, we show that the mean value of $V_{\rm{OGLE-III}}$ is 0.16 mag brighter, on average, than the mean value of $V_{\rm{MACHO}}$. Some of the scatter is due to the fact many OGLE RR Lyrae stars have few measurements in V , which will affect their individual precision.  However, this will not affect their accuracy, and thus not their precision in the mean. A difference in $V_{OGLE-III}$ and $V_{MACHO}$ is therefore unphysical if both $V_{OGLE-III}$ and $V_{MACHO}$ are standard. In particular, a linear regression with respect to reddening yields $E(V_{\rm{OGLE-III}}-V_{\rm{MACHO}})   \propto  (0.064\pm0.024)E(V-I)$, suggesting a 2.67$\sigma$ detection that $V_{\rm{MACHO}}$ is centered at a lower effective wavelength, with an implied difference of $\sim$150 \AA. However, this trend could also be due to blending of fainter sources or a non-linear response in the detectors, as higher $E(V-I)$ also means fainter stars.

In the bottom panel of Figure \ref{Fig:RRmachoogle}, we plot the behavior of $E(V-R)_{\rm{MACHO}} /E(V-I)_{\rm{OGLE-III}}$. A linear regression yields $E(V-R)_{\rm{MACHO}} /E(V-I)_{\rm{OGLE-III}} = (0.544\pm0.027)+(-0.048\pm0.009)(E(V-I)-1.033)$. The mean value of 0.544 is problematic. For a \citet{1989ApJ...345..245C} extinction curve, $E(V-R) /E(V-I)\approx0.42$ for nearly all values of $R_{V}$, but only if $R$ is centered at $\sim$6500 \AA \, as per the Kron-Cousins system \citep{1983PASP...95..480B}. If the bandpass is in fact centered near  $\sim$6900 \AA, as per the Johnson system \citep{1983PASP...95..480B}, then the expected reddening ratio is $E(V-R)_{\rm{MACHO}} /E(V-I)_{\rm{OGLE-III}}\approx0.57$. However, given the uncertainties in $V_{MACHO}$, it is difficult to tell if this difference represents an additional problem in $R_{MACHO}$ or simply the error from $V_{MACHO}$ propagating, through both an uncertain zero-point calibration for the reddening law, and with the assumption of an  intrinsic RR Lyrae color at minimum light of $(V-R)_{\rm{RRab},0}=0.28$. 

Further evidence for errors in the zero points is suggested by fitting for the two reddening values. We find $E(V-R)_{MACHO} = (0.0519\pm0.009) +(0.490\pm0.008)E(V-I)$. The y-intercept  deviates from the origin at the 5.7$\sigma$ level, in spite of the fact that $E(V-R)_{MACHO}$ is expected to be zero when $E(V-I)$ is zero. To have a y-intercept of zero, one would either need to shift the intrinsic colour of the RC by 0.11 mag, to $(V-I)_{RC,0}=0.95$, or adjust the minimum-light color of MACHO RR Lyrae by 0.05 mag, to $(V-R)_{\rm{RRab},0}=0.33$. Both are uncomfortably large changes: The value $(V-I)_{RC,0}=1.06$ was shown to be accurate to a few hundredths of a magnitude in Section \ref{sec:Calibration}, and  \citet{2010AJ....139..415K} measured $(V-R)_{\rm{RRab},0}=0.28 \pm 0.02$ using a sample of local RR Lyrae. Alternatively, composite extinction bias (see Section \ref{subset:composite}) might play a role.



The OGLE-III filters are rigorously calibrated on the Landolt photometric system to an accuracy of a few hundredths of a magnitude \citep{2011AcA....61...83S}. We therefore conclude that there is likely an error in the calibration of the MACHO filters. 

\section{Constraining Fundamental Parameters of Galactic Structure I: The Galactocentric Distance and the Viewing Angle to the Galactic Bulge}
\label{sec:GalacticStructure}
The structure of the inner Galaxy is currently a matter of active research, as the spatial morphology of the bulge is not fully known. There is an X-shaped component at large separations from the plane \citep{2010ApJ...721L..28N,2010ApJ...724.1491M,2011AJ....142...76S,2012ApJ...757L...7L,2012ApJ...756...22N}, evidence for a "long bar'' component at large separations from the minor axis \citep{2005ApJ...630L.149B,2008A&A...491..781C} that may be due to leading ends \citep{2011ApJ...734L..20M}, and an inner structure that is either a secondary, "nuclear bar'' or a viewing effect \citep{2001A&A...379L..44A,2005ApJ...621L.105N,2011A&A...534L..14G,2012ApJ...744L...8G,2012A&A...538A.106R}. There is also evidence that the metal-poor stars are distributed as a classical bulge \citep{2010A&A...519A..77B,2011ApJ...732L..36D,2012A&A...546A..57U}, whereas the data on metal-rich M-giants are  consistent with entirely cold kinematics \citep{2012AJ....143...57K}. The functional dependence of RC brightness on direction has historically been a powerful constraint on bulge properties, responsible for many of the insights mentioned here. It is thus interesting to see what this new dataset of dereddened RC centroids can tell us about the geometry of the bulge. 

\begin{figure}[H]
\begin{center}
\includegraphics[totalheight=0.6\textheight]{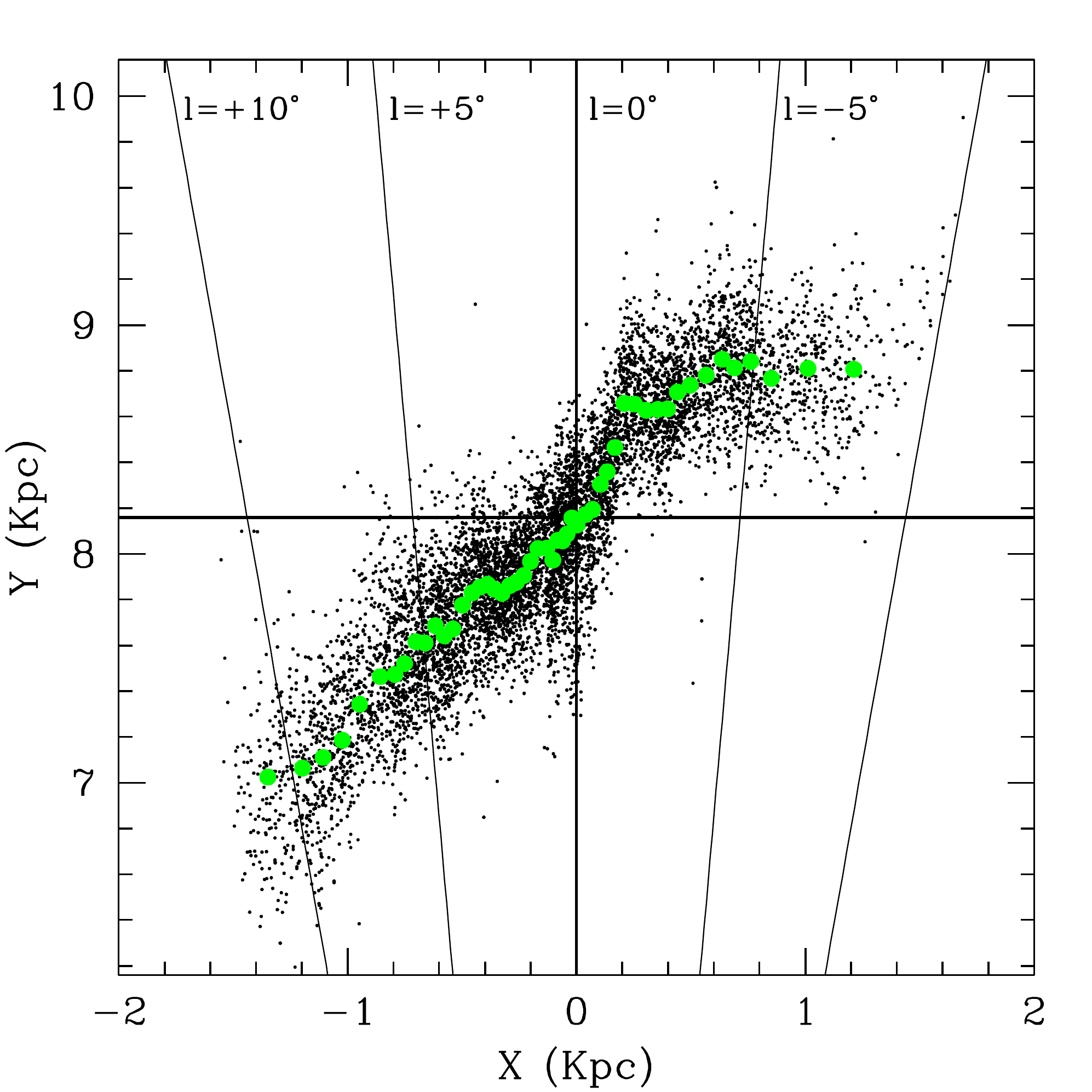}
\end{center}
\caption{Distances to dereddened RC centroids are projected onto a face-on view of the central region of the Milky Way. Black points show the projected location of each measured RC centroid, and green points show 50 binned values.} 
\label{Fig:BarProfile}
\end{figure}

We plot the dereddened apparent magnitudes to the RC centroids measured in this work in Figure \ref{Fig:BarProfile}, where we assume $M_{I,RC}=-0.12$, as per Section \ref{sec:Tuc6791}. We use the same Galactic coordinate system as \citet{1995ApJ...445..716D}. A line fit in the X-Y plane to RC centroids satisfying $|l| \leq 3.0^{\circ}$ yields a Galactocentric distance $R_{0}=8.20$ kpc and an apparent viewing angle of $\alpha=40^{\circ}$, which is a soft upper-bound and consistent with a true viewing angle of $\alpha= 25-27^{\circ}$  (explained below). 

Our estimate of $R_{0}=8.20$  resolves a dissonance in the literature. The extinction map of \citet{2004MNRAS.349..193S}, which assumed that $A_{I}/E(V-I) = dA_{I}/dE(V-I)$, yielded a distance to the Galactic center of $\sim$9 kpc \citep{2007MNRAS.378.1064R,2009A&A...498...95V} when applied to $VI$ photometry of RC stars. That value is too large when contrasted to the geometrically determined distances to the Galactic center of $7.62 \pm 0.32$ kpc \citep{2005ApJ...628..246E}, $8.27 \pm 0.29$ kpc \citep{2012arXiv1207.3079S},  and $8.4 \pm 0.4$ kpc \citep{2008ApJ...689.1044G}. It is also larger than distances derived from studies of the RC in the near-IR \citep{2005MNRAS.358.1309B,2006ApJ...638..839N}. \citet{2007MNRAS.378.1064R} actually shifted their measured values of $I_{RC}$ by 0.3 mag to force the bulge to be centered at $R_{0} = 8$ kpc, whereas \citet{2009A&A...498...95V} commented that ``The origin of this discrepancy is not understood at the moment''. Our distance of 8.2 rather than 8.0 kpc reduces that offset by 0.05 mag, and our corrected calibration of $M_{I,RC}$ further reduces it by 0.14 mag. The different extinction law accounts for the rest. The solution to the non-standard $VI$ extinction toward the Galactic bulge thus relieves a major bottleneck in Galactic bulge studies: there is no peculiar population effect or catastrophic failure of stellar evolution models decalibrating our primary standard candle by a spectacular value of 0.3 mag.

The margin for the population correction of the bulge RC, and thus the error on our estimate $R_{0}=8.20$ kpc, is rapidly shrinking. The spectroscopic metallicity distribution function of the Galactic bulge is now known to an impressive degree of accuracy \citep{2008A&A...486..177Z,2010A&A...513A..35A,2011A&A...534A..80H,2011A&A...533A.134B,2011ApJ...732..108J,2012ApJ...746...59R}. The same is not true of the age and helium abundance of the Galactic bulge. As the uncertainty in these two parameters is only an issue for the more metal-rich stars \citep{2011ApJ...735...37C,2011A&A...533A.134B,2012ApJ...751L..39N}, its integrated effect will not be very large. A lower age or higher helium abundance for the bulge would both require a larger value of $R_{0}$, see Equation (\ref{EQ:colorterms3}).

\begin{figure}[H]
\begin{center}
\includegraphics[totalheight=0.37\textheight]{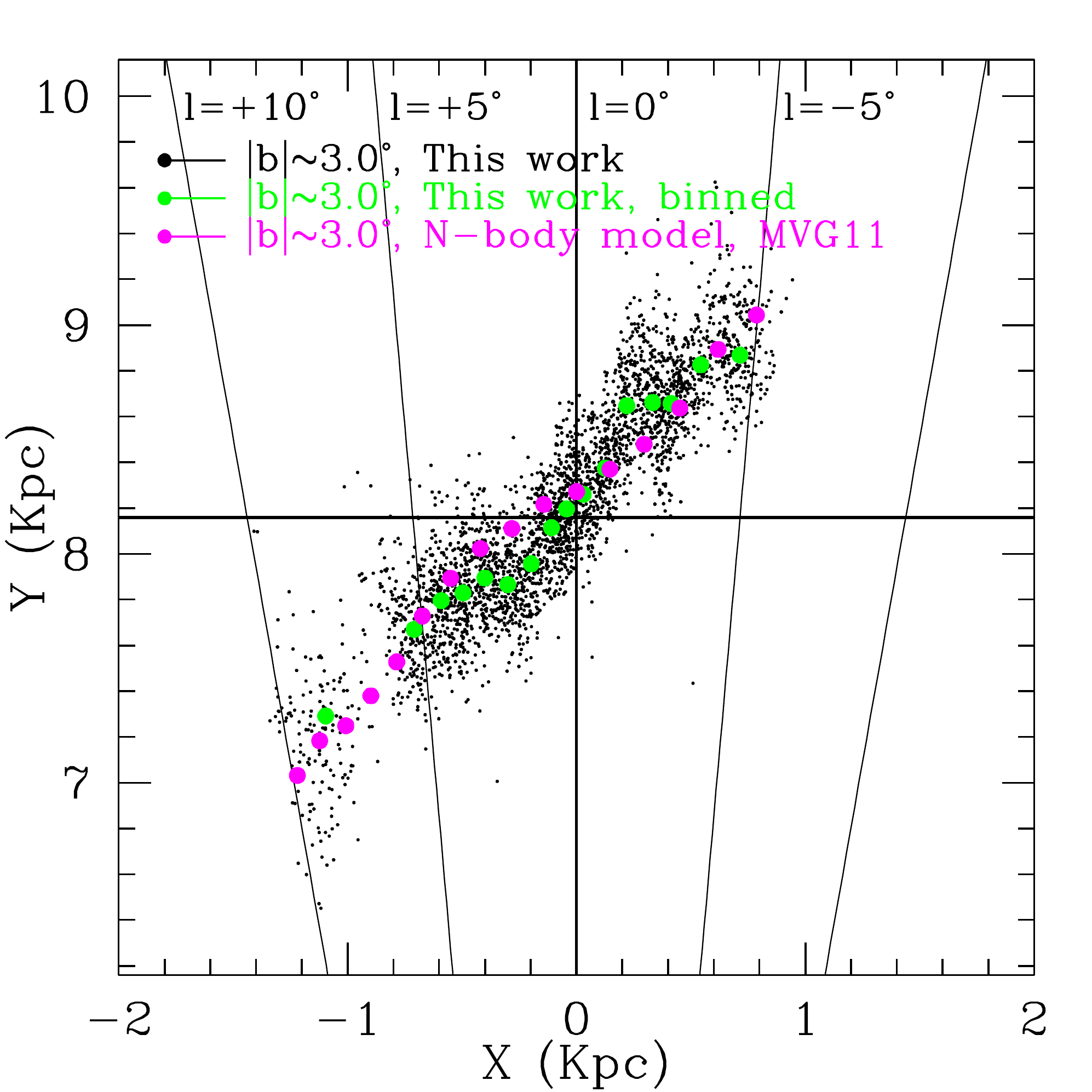}
\end{center}
\caption{Data from this work (black), binned data from this work (green), and predictions (magenta) from the N-body model of \citet{2011ApJ...734L..20M}, for sightlines satisfying $(-10.5^{\circ} \leq l \leq +5.5^{\circ}, 2.5^{\circ} \leq |b| \leq 3.5^{\circ})$. } 
\label{Fig:BarProfile2}
\end{figure}

\begin{figure}[H]
\begin{center}
\includegraphics[totalheight=0.37\textheight]{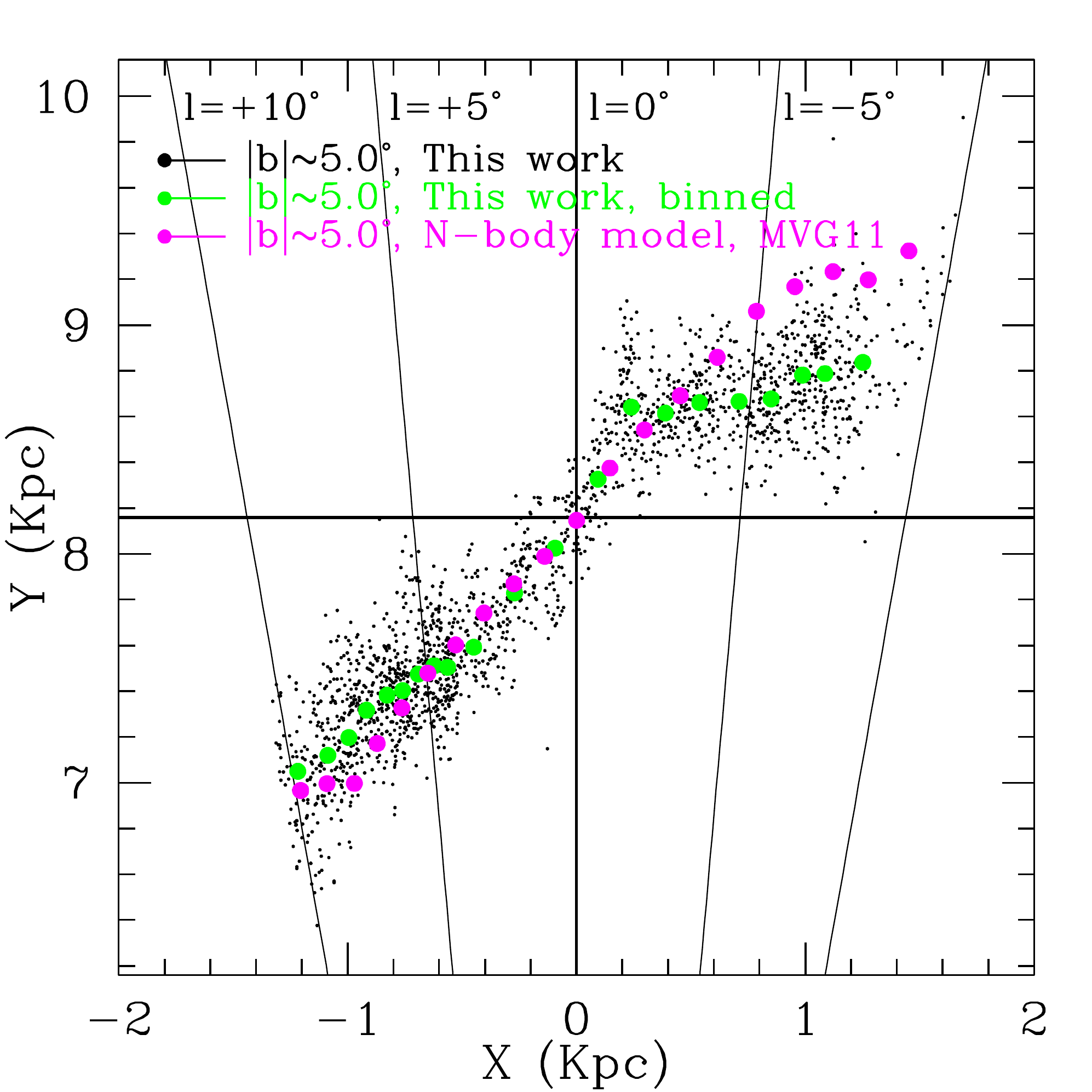}
\end{center}
\caption{Color scheme as in Figure \ref{Fig:BarProfile2}, for sightlines satisfying $(-10.5^{\circ} \leq l \leq +10.5^{\circ}, 2.5^{\circ} \leq |b| \leq 3.5^{\circ})$. } 
\label{Fig:BarProfile3}
\end{figure}

The value of $\alpha=40^{\circ}$ is a soft upper bound. This is because the distance along the plane to the maximum density along the line of sight to a triaxial structure is strictly less than the distance to that structure's major axis for that sightline on the far side, and strictly greater on the near side. The difference can lead to a bias of $\sim$50\% in the inferred viewing angle of the Galactic bulge, and depends on the bulge's axis ratios \citep{1994ApJ...429L..73S,2007A&A...465..825C}. This can also be discerned by investigating viewing effects by means of sophisticated dynamical models. We compare our results to predictions from the N-body model used by \citet{2011ApJ...734L..20M} for two latitudes in Figures \ref{Fig:BarProfile2} and \ref{Fig:BarProfile3}. As their analysis assumes $R=8$ kpc, we have rescaled their distances by 1.025 to be consistent with the distance inferred in this work. Though their assumed viewing angle is $\alpha=25^{\circ}$, their model points line up nearly perfectly with our data points. The apparent angle from the model is $\sim 1^{\circ}$ greater for the data in Figure \ref{Fig:BarProfile2} and $\sim 5.5^{\circ}$ greater in Figure \ref{Fig:BarProfile3}. As the apparent angle is twice as large as the true angle, this suggests a superior match would be attained if the assumed angle of the model was set to $\alpha \sim 27^{\circ}$. There are further discrepancies that warrant further investigation, for example in Figure ref{Fig:BarProfile3}, the predicted and observed data points do not line up well for $X \geq 0.5 kpc$.

\section{The Apparent Magnitude of the Red Clump as an Improved Constraint on Microlensing Events}
\label{sec:MicrolensingConstraint}
A fundamental parameter of microlensing lightcurves is the angular size of the Einstein ring, which is proportional to the square root of the mass of the lens. It is related to observables by:
\begin{equation}
\theta_{E} = \frac{{\theta_{*}}}{\rho},
\end{equation}
where $\rho$ is a term due to finite-source effects that can be directly measured  from well-sampled lightcurves \citep{1994ApJ...421L..71G,1994ApJ...424L..21N,1994ApJ...430..505W}. The angular size of the source, ${\theta_{*}}$, is given by:
\begin{equation}
F = \pi S {\theta_{*}}^2,
\end{equation}
where $F$ is the dereddened flux of the source, and $S$ is the dereddened surface brightness. 

The first step used in inferring the surface brightness is measuring the color difference between the source and the RC, where the intrinsic color of the RC is calibrated by \citet{2011A&A...533A.134B}. A combination of color-color relations \citep{1988PASP..100.1134B} and  empirically-calibrated relations between color and surface brightness \citep{2004A&A...426..297K} can then be used to yield the surface brightness. It is not as straightforward to infer the flux, due to the lack of a calibration for the intrinsic magnitude of the RC and the degeneracy between extinction and distance. The flux has been inferred by assuming a value of $M_{I,RC}$, a distance to the Galactic center, and a correction for the distance due to the Galactic bulge's orientation angle \citep{2004ApJ...603..139Y}.

The results of this work will greatly simplify this process by streamlining the step that may have been the largest source of error: the determination of the dereddened apparent magnitude of the source. The bottom-left panel of Figure \ref{Fig:LongMagMosaic3B} shows that the dereddened apparent magnitude of the RC, $I_{RC,0}$, is a very well-behaved function of longitude, for which we estimated an intrinsic scatter of $\sim$0.04 mag in Section \ref{sec:AIdetermination}. We summarize the results in Table \ref{table:MicrolensingHelp}. The need to assume a distance to the Galactic center, a value of $M_{I,RC}$, and a correction due to the Galactic bulge's orientation is no more: one can now simply measure the brightness difference between the RC and the source, and then read off the value $I_{RC,0}$ as a function of $l$ from Table \ref{table:MicrolensingHelp}. 

\begin{table}[H]
\caption{Dereddened magnitude of the RC, $I_{RC,0}$, as a function of Galactic longitude. The 1$-{\sigma}$ measurement error due to metallicity gradients and latitudinal projection effects is no more than 0.04 mag. \newline}
\centering 
\begin{tabular}{ll}
	\hline \hline
$l$ (deg) & $I_{RC,0}$ \\
	\hline \hline \hline
-9 & 14.662 \\ 
-8 & 14.624 \\ 
-7 & 14.620 \\ 
-6 & 14.619 \\ 
-5 & 14.616 \\ 
-4 & 14.605 \\ 
-3 & 14.589 \\ 
-2 & 14.554 \\ 
-1 & 14.503 \\ 
\,\,0 & 14.443 \\ 
\,\,1 & 14.396 \\ 
\,\,2 & 14.373 \\ 
\,\,3 & 14.350 \\ 
\,\,4 & 14.329 \\ 
\,\,5 & 14.303 \\ 
\,\,6 & 14.277 \\ 
\,\,7 & 14.245 \\ 
\,\,8 & 14.210 \\ 
\,\,9 & 14.177 \\ 
\,\,10 & 14.147 \\ 
\,\,11 & 14.121 \\ 
	\hline
\end{tabular}
\label{table:MicrolensingHelp}
\end{table}



\newpage

\section{Constraining Fundamental Parameters of Galactic Structure II: \\ The Stellar Mass of the Galactic Bulge}
\label{sec:GalacticStructure2}
The precise stellar mass of the Galactic bulge remains a mystery. A widely quoted value is $1.3 {\times}10^{10} M_{\odot}$ from the work of \cite{1995ApJ...445..716D}, who converted the bolometric luminosity of the bulge into a number of RG stars, and then obtained a total mass by integrating a Salpeter initial mass function \citep{1955ApJ...121..161S} down to $M=0.1M_{\odot}$. In contrast, \citet{1995ApJ...444L..89B} estimated $(1.7-2.8) {\times}10^{10} M_{\odot}$ by using the tensor virial theorem, and \citet{1996MNRAS.283.1197Z} estimated $M \geq 2.0 {\times}10^{10} M_{\odot}$ based on the frequency of gravitational microlensing events toward the bulge. \citet{2012A&A...538A.106R} estimated $1.1 {\times}10^{10} M_{\odot}$ by fitting 2MASS data to various parametric models.

A baryonic mass for the Galactic bulge would constrain models of Milky Way assembly, both in its own right and by anchoring the mass scale of the Galaxy as a whole. It is thus interesting to ask how well this dataset could eventually contribute. Most RC studies have been focused on tracing distances, but the Paczynski-Stanek equation also allows one to trace star counts, as $N_{RC}$ is a free-parameter. We thus provide a constraint on the stellar mass of the Galactic bulge by estimating the number of RC stars.  

\begin{figure}[H]
\begin{center}
\includegraphics[totalheight=0.8\textheight]{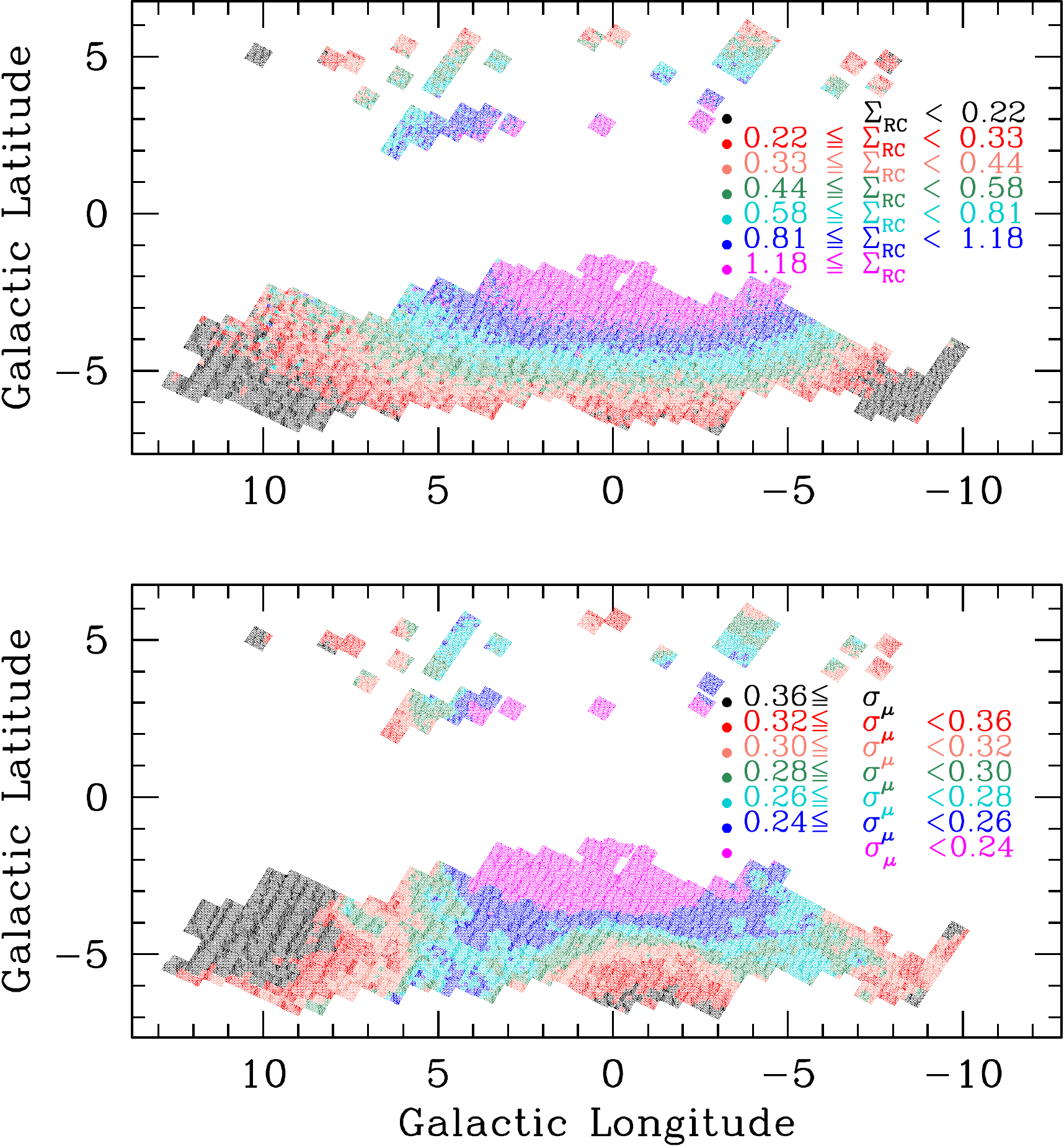}
\end{center}
\caption{TOP: Surface density of RC stars toward the Galactic bulge ${\Sigma}_{RC}$, as a function of direction. Values are normalized to the surface density toward Baade's window $(l=1^{\circ},b=-3.9^{\circ})$ of 50,100 RC stars deg$^{-2}$. BOTTOM: Distance modulus dispersion of bulge RC stars as a function of direction, after application of a 20$\arcmin$ smoothing..   } 
\label{Fig:Mapper17D4Paper4Maps3}
\end{figure}

\begin{figure}[H]
\begin{center}
\includegraphics[totalheight=0.78\textheight]{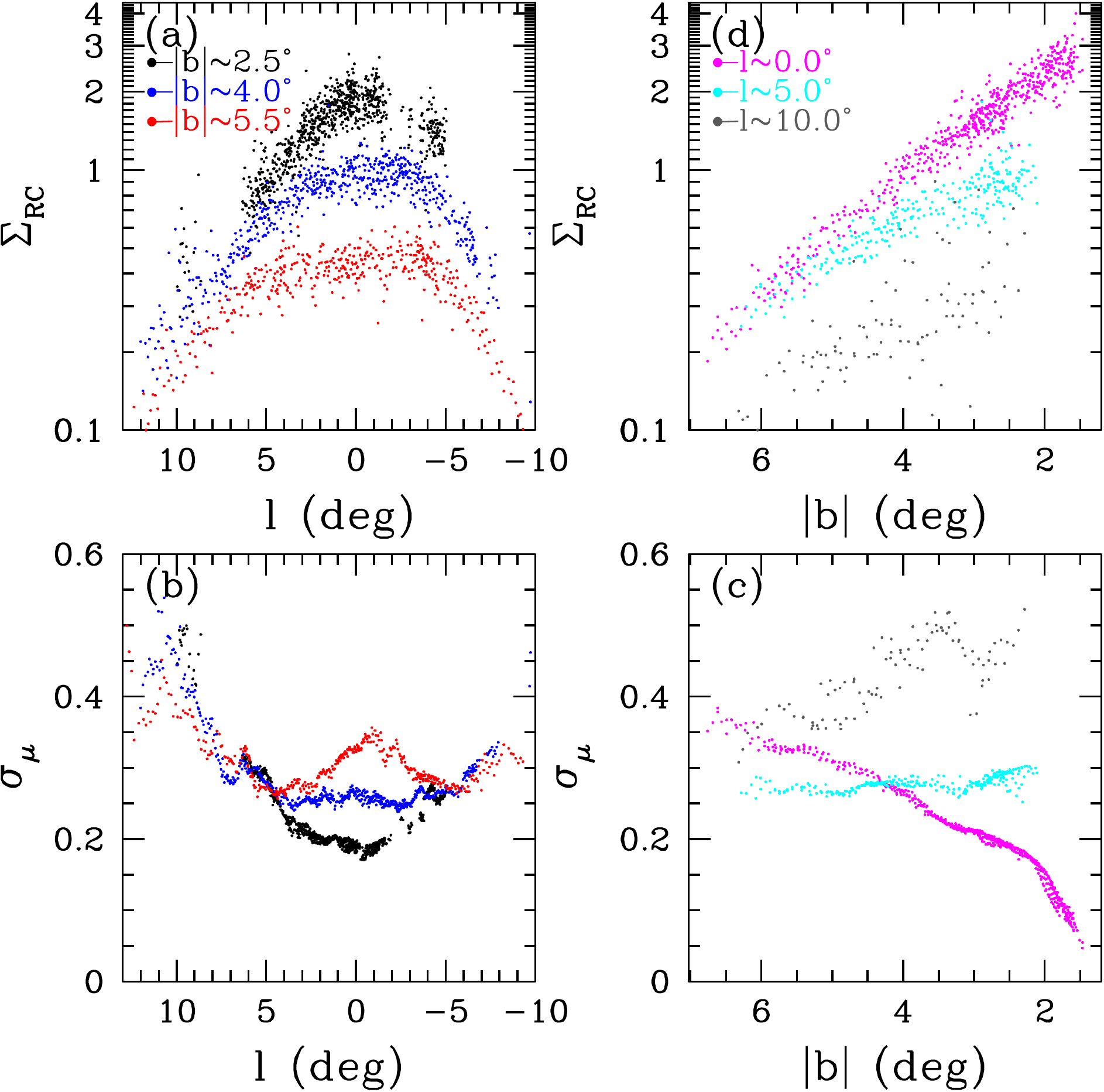}
\end{center}
\caption{The number density of RC stars (normalized to that of 50,100 deg$^{-2}$ toward Baade's window) and dispersion in distance modulus (with 20$\arcmin$ smoothing) as a function of direction.\textbf{ Panel (a)} ${\Sigma}_{RC}$ as a function of longitude for $|b|\sim 2.5^{\circ}$ (black), $|b|\sim 4.0^{\circ}$ (blue), and  $|b|\sim 5.5^{\circ}$ (red). \textbf{Panel (b)} $\sigma_{\mu}$ as a function of longitude for three latitudinal stripes, same color scheme as panel (a). \textbf{Panel (c)} $\sigma_{\mu}$ as a function of latitude for  $l\sim 0.0^{\circ}$ (magenta), $l\sim 5.0^{\circ}$ (cyan), and  $l\sim 10.0^{\circ}$ (grey). \textbf{Panel (d)} ${\Sigma}_{RC}$ as a function of latitude for three longitudinal stripes, same color scheme as panel (c).  } 
\label{Fig:NumbersWidth}
\end{figure}

We plot the surface density of RC stars, ${\Sigma}_{RC}$, in the top panel of Figure \ref{Fig:Mapper17D4Paper4Maps3}, where we normalize to the density of 50,100 RC stars deg$^{-2}$ observed toward Baade's window. We also show the scatter of ${\Sigma}_{RC}$ for latitudinal and longitudinal stripes on the top-left and top-right panels of Figure \ref{Fig:NumbersWidth}, respectively. The top-left panel demonstrates that $\Sigma_{RC}$ is maximized near $l=0^{\circ}$ for all latitudinal stripes, however, the sharp peak turns to a plateau at large separations from the plane. The top-right panel shows the expected result that number counts increase monotonically with decreased separation from the plane at fixed longitude.


We measure $2.94 {\times} 10^{6}$ RC stars in OGLE-III over a viewing area of $90.25 \rm{\,deg}^{2}$ for $\sim$9,000 sightlines deemed reliable. We extrapolate this across the bulge using the ``G1'' model of \citet{1995ApJ...445..716D}, assuming a viewing angle $\alpha=25^{\circ}$, a corotation radius of 4.0 kpc, and a galactocentric distance of $R_{0}=8.20$ kpc. We find a best-fit predicted RC population for the bulge of $14.6 {\times} 10^{6}$, with output axial ratios of $X_{0}:Y_{0}:Z_{0}=1:0.41:0.29$. In contrast, fixing $\alpha=40^{\circ}$ yields $13.9 {\times} 10^{6}$ RC stars with axis ratios of $X_{0}:Y_{0}:Z_{0}=1:0.40:0.38$. The X-shaped component should not be a significant source of uncertainty -- \citet{2012ApJ...757L...7L} estimate that it should contribute only $\sim$7\% of the total stellar mass of the bulge.

Converting this population number into a stellar mass requires assumptions as to the stellar lifetime of the RC and the IMF of the bulge stellar population. For an old, metal-rich population, \citet{2011arXiv1109.2118N} used stellar models to compute that the initial mass of stars found on the RG branch (and thus at the helium flash) is predicted to be:
\begin{equation}
\log (M/M_{\odot}) = 0.026 + 0.126{\times}\rm{[M/H]} -0.276{\times}\log(t/(10\rm{\,Gyr}))-0.937{\times}(Y-0.27).
\label{EQ:RGmass}
\end{equation}
The lifetime of the helium-burning phase is approximately 100 million years \citep{2000A&A...361.1023S}. Assuming a mean metallicity of [M/H]$=+0.16$ \citep{2011A&A...534A..80H}, a mean age of $t=10$ Gyr \citep{2011ApJ...735...37C,2011A&A...533A.134B} and a mean initial helium abundance of Y=0.33 \citep{2011ApJ...730..118N,2012ApJ...751L..39N} leads to the estimate that RC stars originate from the initial mass range $(0.9744 M_{\odot} \leq M \leq 0.9771 M_{\odot})$. We assume a Salpeter IMF  \citep{1955ApJ...121..161S} for the bulge over the range 0.1 $M_{\odot} \leq M_{\rm{Initial}} \leq 100 M_{\odot}$, that 90\% of bulge stars that complete their stellar evolution through the helium-burning phase (discussed below), and the same remnant mass function as \citet{2000ApJ...535..928G}, whose assumptions are based on the results of \citet{1995ApJ...443..735B} and \citet{1999ApJ...512..288T}. We end up with an estimated stellar mass for the Galactic bulge of $2.3{\times}10^{10} M_{\odot}$. 


There are several uncertainties in this estimate, which are tractable in principle. The use of a triaxial model from \citet{1995ApJ...445..716D} is problematic -- these triaxial ellipsoid models are now ruled out by the data, for reasons listed at the top of Section \ref{sec:GalacticStructure}. However, our purpose in this section is merely to provide an estimate. We publish all our measured RC parameters and it will be straightforward to reproduce our calculations once more sophisticated structural models are available.  An additional underestimate arises from our assumption that 90\% of stars that ``should'' have ended up on the RC did end up on the RC. In reality, many bulge stars either end up on the extreme blue horizontal branch or skip the helium-burning phase altogether. The \textit{HST} photometry used by \citet{2011ApJ...735...37C} shows that at least $\sim$3\% of bulge horizontal branch stars are not RC stars, and the fraction could be found to be higher if deeper photometry is obtained. Moreover, 10\% of field white dwarfs have masses lower than the helium-ignition limit \citep{2005ApJS..156...47L}, some of which are not in binaries \citep{2011ApJ...730...67B}. These would not even show up on the blue horizontal branch, but should be incorporated into a complete model. Our estimate assumes that the fraction of bulge stars that skip the helium-burning phase can be inferred from the fraction of local white dwarfs with masses smaller than that of the helium-burning limit. The third source of error is in our evolutionary assumptions for Equation \ref{EQ:RGmass}: a higher stellar mass for the bulge would result from assuming a lower mean age, a lower helium abundance, or a higher mean metallicity. 

Finally, the assumption of a Salpeter IMF is likely the most significant source of error: our estimate of the total stellar mass would drop by one third if we assumed the same IMF of \citet{2000ApJ...530..418Z}.  \citet{2000ApJ...530..418Z} used \textit{HST} observations of the bulge luminosity function to estimate $\alpha=-1.33 \pm 0.07$, which is very bottom-light relative to the value of $\alpha=-2.35$ for a Salpeter IMF.  \citet{2008A&A...480..723C} used the duration distribution of Galactic gravitational microlensing events toward the bulge to estimate $\alpha=-1.70 \pm 0.5$. On the other hand, \citet{2012arXiv1205.6473C}  have argued that the properties of M-dwarf spectral features in the integrated light of field galaxies implies that stellar systems which are metal-rich and $\alpha$-enhanced have bottom-heavy IMFs. Their analysis is not consistent with that of \citet{2000ApJ...530..418Z} and \citet{2008A&A...480..723C}, and thus further investigation is warranted. The spectacular Galactic bulge luminosity functions of \citet{2010ApJ...725L..19B}, measured with \textit{HST}, may provide a path to better determining the Galactic bulge IMF, and thus the total bulge stellar mass.

\section{Constraining Fundamental Parameters of Galactic Structure III: \\ The Geometrical Thickness of the Galactic Bulge}
\label{sec:GalacticStructure3}
The thickness of the Galactic bulge (ratio of minor to major axis) is a very sensitive probe of the environmental conditions in which the Milky Way's bulge formed and evolved, and disagreements as to the thickness have been a catalyst to significant disagreements as to the nature of the Milky Way's bulge \citep{2011arXiv1106.0260L}. Further, detailed investigations of the Galactic bulge luminosity function require not only the first moment of the distance distribution, but also the second moment as well. 

We estimate the geometrical thickness of the Galactic bulge in units of distance modulus, $\sigma_{\mu}$, as follows:
\begin{equation}
{\sigma}_{\mu}^2 = \sigma_{I,RC}^2 - \sigma_{I,RC,0}^2 - R_{I}^2{\times}\biggl(\sigma_{(V-I)}^2 - \sigma_{(V-I),RC,0}^2  \biggl),
\label{EQ:Thickness}
\end{equation}
where $\sigma_{I,RC}$ is the measured brightness dispersion of the RC, $\sigma_{I,RC,0} = 0.09$ is the estimated intrinsic magnitude dispersion of the RC in $I$, and $R_{I}^2{\times}(\sigma_{(V-I)}^2 - \sigma_{(V-I),RC,0}^2  )$ is the differential extinction component, with $R_{I}=A_{I}/E(V-I)$ measured as per Equation \ref{EQ:FinalReddeningLaw}. Due to the fact the measurement error on $\sigma_{I,RC}$ is large relative to the variations, we smooth the measurements by replacing each value of ${\sigma}_{\mu}^2$ with the mean of all the values of  ${\sigma}_{\mu}^2$ located within 20$\arcmin$. The values of ${\sigma}_{\mu}$ as a function of direction are plotted in the bottom panel of Figure \ref{Fig:Mapper17D4Paper4Maps3}. We also show the scatter of ${\sigma}_{\mu}$ for latitudinal and longitudinal stripes on the bottom-left and bottom-right panels of Figure \ref{Fig:NumbersWidth}, respectively. The bottom-left panel shows that for sightlines close to the plane, ${\sigma}_{\mu}$ is minimized near $l=0^{\circ}$, the expected behavior for a triaxial ellipsoid. Conversely, for sightlines further from the plane, ${\sigma}_{\mu}$ is minimized near $l= \pm 5^{\circ}$. \cite{2010ApJ...721L..28N} used the increase in ${\sigma}_{I,RC}$ toward $(l \approx 0^{\circ}, |b| \gtrsim 4.75^{\circ})$ to infer the existence of the double RC. This increased geometric dispersion with separation from the plane for sightlines along the minor axis is matched by N-body models with an X-shaped bulge \citep{2012ApJ...757L...7L,2012ApJ...756...22N}. Further from the minor axis, we suspect that the high values of ${\sigma}_{\mu}$ for $l \gtrsim 8^{\circ}$ are due to disk contamination. The bottom-right panel of Figure \ref{Fig:NumbersWidth} shows that $\sigma_{\mu}$ is correlated with separation from the plane along the minor axis (as would be expected of a bulge), and becomes \textit{anti-correlated } with separation from the plane at large separations from the minor axis. The latter is the expected behavior if disk contamination is high for $l=10^{\circ}$: sightlines closer to the plane will probe more of the disk and thus a larger range of distances. We note that the very low values of $\sigma_{\mu}$ close to $(l,b)=(0^{\circ},-2^{\circ})$ are more sensitive to possible errors in our zero-point calibrations.

\begin{table}[H]
\caption{Values of the mean distance modulus (assuming $M_{I,RC}=-0.12$), distance modulus dispersion, reddening, differential reddening, and reddening law (Equation \ref{EQ:FinalReddeningLaw})  for four sightlines that are the subject of intensive \textit{HST} observations \citep{2006Natur.443..534S,2010ApJ...725L..19B,2011ApJ...735...37C}. $E(V-I)$ and $\sigma_{E(V-I)}$ for these sightlines were measured using circles centered on these field centers with radii of 3$\arcmin$.  \newline}
\centering 
\begin{tabular}{|c|c|c|c|c|c|c|c|rr}
	\hline \hline
Field Name       & $l$ (deg)&$b$ (deg) & ${\mu}$& $\sigma_{\mu}$  & $E(V-I)$ & $\sigma_{E(V-I)}$ & $R_{JKVI}$ \\
	\hline \hline \hline
Stanek's Window   & $+0.25$  & $-2.15$ &  14.53 &  0.17    & 1.04  & 0.08   & 0.375    \\ \hline
SWEEPS           & $+1.26$  & $-2.65$ & 14.52  & 0.20     & 0.79  & 0.10   & 0.362  \\ \hline
Baade's Window   & $+1.06$  & $-3.81$ &  14.54 & 0.24     & 0.67 & 0.04  & 0.351    \\ \hline
OGLE 29          & $-6.75$  & $-4.72$ & 14.76  & 0.29     & 0.67 & 0.00   & 0.321  \\ \hline
	\hline
\end{tabular}
\label{table:GeometricDispersions}
\end{table}

The geometric dispersion (and other parameters) toward four bulge fields of high scientific interest, including  Baade's window and Stanek's window \citep{1998astro.ph..2307S}, are listed along with other parameters in Table \ref{table:GeometricDispersions}. We note that the geometric dispersions measured here apply only to the stellar population adequately traced by the RC. Metal-poor bulge stars are kinematically hotter \citep{2010A&A...519A..77B,2011ApJ...732L..36D,2012ApJ...750..169P,2012A&A...546A..57U}, and will thus have a larger geometric dispersion. 

\section{Summary of Data}
\label{Sec:DataSummary}
The results of this work are available for download on the OGLE webpage \footnote{http://ogle.astrouw.edu.pl/}. We briefly summarize the format here. The first table is intended for observers, of which we show a cropped version as Table \ref{table:ForObservers}. It includes values of the reddening, extinction, differential reddening, reddening law, mean distance modulus, and distance modulus dispersion for each sightline. The final reddening map contains $\sim$19,000 observations on which to base its interpolations, higher than the $\sim$9,000 used for calibrations in this work. The denser grid uses the prior grid as a soft prior and was computed to increase the accuracy of interpolation. The OGLE webpage includes a GUI to facilitate retrieval of these values for a given coordinate. 

The determinations of $\mu$, $\sigma_{\mu}^2$, $N_{RC}$ and full error matrix thereof for $\sim$9,000 sightlines investigated in this work are also available in a separate table. We show a cropped version as Table \ref{table:ForModellers}.

\begin{table}[H]
\caption{Coordinates, extinction, reddening, differential reddening, mean distance modulus, distance modulus dispersion, reddening law and Quality Flag for each of $\sim$9,000 sightlines studied in this work. A 20$\arcmin$ smoothing is applied to the values of the mean and dispersion of the distance modulus distribution. A Flag of ``0'' means a reliable measurement. \newline}
\centering 
\begin{tabular}{ccccccccc}
	\hline \hline
$l$ & $b$ & $A_{I}$ & $E(V-I)$ & $\sigma_{E(V-I)}$ & $\mu$ & $\sigma_{\mu}$ & $R_{JKVI}$ & Flag \\
	\hline \hline 
-10.06 & -4.32 & 0.89 & 0.78 & 0.07 & 14.87 & 0.34 & 0.29 & 0 \\ \hline
-9.94  & -4.24 & 1.02 & 0.87 & 0.09 & 14.89 & 0.39 & 0.32 & 0 \\ \hline
-9.91  & -4.57 & 0.85 & 0.72 & 0    & 14.86 & 0.35 & 0.31 &  0 \\ \hline
	\hline
\end{tabular}
\label{table:ForObservers}
\end{table}

\begin{table}[H]
\caption{Structural parameters for sightlines deemed reliable, with full error matrix.  The symbol ``$S$'' denotes the 1-$\sigma$ measurement error on the variable, as opposed to the more standard ``$\sigma$'' to avoid confusion with $\sigma_{\mu}$, and $C$ denotes the correlation. Due to measurement errors, approximately 2\% of sightlines have a best-fit negative variance to the distance modulus distribution. \newline}
\centering 
\begin{tabular}{ccccccccccc}
	\hline \hline
$l$ & $b$ & $\mu$ & $\sigma_{\mu}^2$ & $N_{RC}$ & $S{\mu}$ & $S{\sigma_{\mu}^2}$  & $S{N_{RC}}$ & $C_{(\mu,\sigma_{\mu}^2)}$  & $C_{(\mu, N_{RC})}$  & $C_{(\sigma_{\mu}^2,N_{RC})}$ \\
	\hline \hline 
-10.06 & -4.32 & 14.86 & 0.13 & 287 & 0.07 & 0.04 &  50 & 0.14 & 0.11 & 0.68 \\ \hline
-9.94 & -4.24 & 14.84 & 0.13 & 271 & 0.07 & 0.05 &  54 & 0.22 & 0.23 & 0.72 \\ \hline
-9.91 & -4.57 & 14.92 & 0.20 & 394 & 0.07 & 0.06 &  74 & 0.33 & 0.37 & 0.82 \\ \hline
	\hline
\end{tabular}
\label{table:ForModellers}
\end{table}

\section{Discussion and Conclusion}
\label{sec:Discussion}
Our solution to the observational problem of the non-standard $VI$ extinction toward the inner Galaxy mitigates what has been one of the dominant sources of uncertainty in studies of the Galactic bulge. The extinction law is on average steeper in the optical with $\sim$30\% variations superimposed, and is well-fit by the relation $A_{I} = 0.7465{\times}E(V-I) + 1.3700{\times}E(J-K_{s})$. The residuals to the extinction fit is now reduced to no more than 0.06 mag, and the estimate of $R_{0} = 8.20$ kpc is consistent with there being no bias in our fit to the extinction law.  In the course of making these measurements, we have also measured that differential reddening averages ($\sim$9\% of total reddening for small fields), and the intrinsic luminosity parameters for the bulge RC. These will be of use to future bulge studies. 

The mean value of $A_{I}/E(V-I)=1.217$ suggests $R_{V}=2.5$, the mean value of $R_{JKVI} = 0.3433$ suggests $R_{V}=2.6$, and thus both the measurements investigated here are consistent with an $R_{V} \approx 2.5$ extinction curve. Measuring the extinction curve in other bandpasses could potentially have major implications for cosmology. Our inferred extinction curve is consistent with the values of $R_{V} \approx 2.5$  inferred in studies of the extinction toward extragalactic type Ia SNe by \citet{2010A&A...523A...7G} and \citet{2011A&A...529L...4C}. In particular, the hierarchical Bayesian analysis of \citet{2011ApJ...731..120M} found that the extinction law toward SNe Ia went as $R_{V}=2.5-2.9$ for $A_{V} \leq 0.4$, and steepened at higher extinctions. \citet{1999ApJ...523..617F} also reported a range in the extinction laws of 23 lensed galaxies of $1.5 \leq R_{V} \leq 7.2$. Since none of the combined 51 measurements of \citet{1989ApJ...345..245C} and \citet{1994ApJ...422..158O} reach such low values of $R_{V}$, there is no reason to expect that extrapolation of this empirical law will behave adequately in domains that lie well beyond its calibration. Indeed, though the values of $A_{I}/E(V-I)=1.217$ and $R_{JKVI} = 0.3433$ are consistent with each other, the change in  $A_{I}/E(V-I)$ as $R_{JKVI}$ changes does not go at the rate predicted by \citet{1989ApJ...345..245C} and \citet{1994ApJ...422..158O}, as shown in Figure \ref{Fig:LongMagMosaic3B}. This domain of the extinction law therefore warrants further investigation. Some constraints could be extracted by combining our results with the recent study of \citet{2012ApJS..201...35N}, who measured reddening values for the color $([3.6{\mu}] - [4.5{\mu}])$.

Our measurements of the number counts, brightness dispersion, mean brightness and full error matrix thereof for $\sim$9,000 RC centroids toward the bulge may be one of the most potent means for constraining the structural parameters of the Galactic bulge. In Sections \ref{sec:GalacticStructure} and \ref{sec:GalacticStructure2}, we have sketched how these data could be used to constrain the morphology and mass of the bulge, without going to the full formalism employed by \citet{1995ApJ...445..716D}, \citet{1997ApJ...477..163S} and \citet{2007MNRAS.378.1064R}. Moreover, as the data have improved, it is now time for the models to improve as well. The use of N-body models by \citet{2007MNRAS.378.1165R}, \citet{2010ApJ...720L..72S}, \citet{2011ApJ...734L..20M}, \citet{2012ApJ...756...22N} and \citet{2012ApJS..201...35N} are encouraging steps in that direction.

\acknowledgments
We thank Mathew Penny, Shude Mao, Daniel Majaess, David Bennett, Andrea Kunder and Krzysztof Stanek for helpful discussions. 

We also thank the anonymous referee for a thorough, detailed, and impactful commentary on our manuscript.

DMN was primarily supported by the NSERC grant PGSD3-403304-2011. DMN and AG were partially supported by the NSF grant AST-1103471. 

The OGLE project has received funding from the European Research Council
under the European Community's Seventh Framework Programme
(FP7/2007-2013) / ERC grant agreement no. 246678 to AU. 

This work has made use of BaSTI web tools.

This publication makes use of data products from the Two Micron All Sky Survey, which is a joint project of the University of Massachusetts and the Infrared Processing and Analysis Center/California Institute of Technology, funded by the National Aeronautics and Space Administration and the National Science Foundation.

\newpage

\appendix



\end{document}